\documentclass[floatfix,twocolumn,showpacs,preprintnumbers,amsmath,amssymb,prr,superscriptaddress,longbibliography, nofootinbib]{revtex4-1}

\usepackage{color}
\usepackage[usenames,dvipsnames,svgnames,table]{xcolor}
\usepackage[colorlinks=true,linkcolor=blue,urlcolor=blue,citecolor=blue]{hyperref}
\usepackage{url}
\usepackage{mathtools}
\usepackage{graphicx}
\usepackage{dcolumn}
\usepackage{array}
\usepackage{lipsum}
\usepackage{bm}
\usepackage{subfigure}
\usepackage{amssymb}
\usepackage{multirow}
\usepackage{tabularx}
\usepackage{amsmath}
\usepackage{braket}
\usepackage{csquotes}
\graphicspath{{plots/}}
 \usepackage{lipsum}
\usepackage{mathrsfs}
\usepackage{MnSymbol}
\usepackage{extarrows}

\newcommand{\beq}{\begin{equation}}
\newcommand{\eeq}{\end{equation}}
\newcommand{\bea}{\begin{eqnarray}}
\newcommand{\eea}{\end{eqnarray}}

\renewcommand{\vec}[1]{\mathbf{#1}}

\begin{document}
\title{ 
Generalized density functional theory framework for the non-linear density response of quantum many-body systems
}

\author{Zhandos~A.~Moldabekov}
\email{z.moldabekov@hzdr.de}

\affiliation{Institute of Radiation Physics, Helmholtz-Zentrum Dresden-Rossendorf (HZDR), D-01328 Dresden, Germany}

\author{Cheng Ma}
\affiliation{Key Laboratory of Material Simulation Methods $\&$ Software of Ministry of Education, College of Physics, Jilin University, Changchun 130012, China. }
\affiliation{State Key Lab of High Pressure and Superhard Materials, College of Physics, Jilin University, Changchun 130012, China. }

\author{Xuecheng Shao}
\affiliation{Key Laboratory of Material Simulation Methods $\&$ Software of Ministry of Education, College of Physics, Jilin University, Changchun 130012, China. }
\affiliation{State Key Lab of High Pressure and Superhard Materials, College of Physics, Jilin University, Changchun 130012, China. }
\affiliation{International Center of Future Science, Jilin University, Changchun 130012, China. }

\author{Sebastian Schwalbe}
\affiliation{Institute of Radiation Physics, Helmholtz-Zentrum Dresden-Rossendorf (HZDR), D-01328 Dresden, Germany}

\author{Pontus Svensson}

\affiliation{Institute of Radiation Physics, Helmholtz-Zentrum Dresden-Rossendorf (HZDR), D-01328 Dresden, Germany}

\author{Panagiotis Tolias}
\affiliation{Royal Institute of Technology (KTH) Stockholm, SE-100 44 Stockholm, Sweden}

\author{Jan Vorberger}
\affiliation{Institute of Radiation Physics, Helmholtz-Zentrum Dresden-Rossendorf (HZDR), D-01328 Dresden, Germany}

\author{Tobias Dornheim}

\affiliation{Institute of Radiation Physics, Helmholtz-Zentrum Dresden-Rossendorf (HZDR), D-01328 Dresden, Germany}

\affiliation{Center for Advanced Systems Understanding (CASUS) at Helmholtz-Zentrum Dresden-Rossendorf (HZDR), D-02826 G\"orlitz, Germany}

{\def\thefootnote{}\footnotetext{To the memory of Dr. Travis Edwin Sjostrom.}} 

\begin{abstract}
A density functional theory (DFT) framework is presented that links functional derivatives of free-energy functionals to non-linear static density response functions in quantum many-body systems. Within this framework, explicit expressions are derived for various higher-order response functions of systems that are homogeneous on average, including the first theoretical result for the cubic response at the first harmonic $\chi_0^{(1,3)}(\vec{q})$. Specifically, our framework includes hitherto neglected mode-coupling effects that are important for the non-linear density response even in the presence of a single harmonic perturbation.
We compare these predictions for $\chi_0^{(1,3)}(\vec{q})$ to new Kohn-Sham DFT simulations, leading to excellent agreement between theory and numerical results. Exact analytical expressions are also obtained for the long-wavelength limits of the ideal quadratic and cubic response functions. Particular emphasis is placed on the connections between the third- and fourth-order functional derivatives of the non-interacting free-energy functional $F_s[n]$ and the ideal quadratic and cubic response functions of the uniform electron gas, respectively. These relations provide exact constraints that may prove useful for the future construction of improved approximations to $F_s[n]$, in particular for warm dense matter applications at finite temperatures.
Here, we use this framework to assess several commonly employed approximations to $F_s[n]$ through orbital-free DFT simulations of the harmonically perturbed ideal electron gas. The results are compared with Kohn-Sham DFT calculations across temperatures ranging from the ground state to the warm dense regime. Additionally, we analyze in detail the temperature- and wavenumber-dependent non-monotonic behavior of the ideal quadratic and cubic response functions.
\end{abstract}

\maketitle

\section{Introduction}

Density functional theory (DFT) is a powerful tool employed in both quantum and classical many-particle physics \cite{Jones_RMP_2015, hansen2013theory}. It is widely used to understand and simulate materials across various phases. The relationship between the functional derivatives of different (free) energy functionals and density response functions connects DFT with other many-body theories, such as Green's functions and, more broadly, kinetic theory~\cite{Vorberger2025,Dornheim_JPSJ_2021,Dornheim_review}. These approaches complement each other, enabling analysis from multiple viewpoints and a deeper understanding of the physics of many-body systems.

The connection between DFT and linear response theory is well established~\cite{marques2012fundamentals,Gross_PRL1985}. In contrast, non-linear response theory within the DFT framework is much less explored~\cite{Moldabekov_jctc_2022}. 
Just as linear response theory has been crucial in advancing approximations for exchange-correlation (XC) functionals~\cite{pribram,Patrick_JCP_2015}, as well as non-interacting free-energy and kinetic-energy functionals in density functional theory (DFT)~\cite{Wenhui_Chemical_Reviews}, non-linear response theory can significantly enhance the quality of models and approximations in DFT. In conceptual DFT, non-linear response properties of electrons are used to understand and characterize chemical bonding \cite{B717671F, Oviedo_jctc_2016, Parr_1983}.
Additionally, as demonstrated here, the DFT framework enables us to identify general forms of solutions for non-linear response functions involving a hitherto unexplored and erroneously neglected mode-coupling mechanism that is important even in the presence of a single external perturbation. This can serve as a useful guide for more tedious theoretical approaches, such as the method of Green's functions and quantum kinetic theory \cite{Vorberger2025, Tolias_2023, William_Carter2019}. Consequently, understanding the relationship between various types of energy functionals and non-linear response functions is essential for both the development of theory and the enhancement of DFT-based simulation methods.

In this work, we explore a general framework for DFT that links the functional derivatives of free-energy functionals to non-linear density response functions. This approach offers a straightforward way to identify how the coupling between different modes contributes to the non-linear response in different situations. We demonstrate this by providing the solution for the cubic response at the first harmonic $\chi_0^{(1,3)}(\vec{q})$ (i.e., the leading non-linear term at the original perturbation~\cite{moroni, Dornheim_PRR_2021}) in terms of mode coupling between density perturbations at the first and second harmonics. This solution, long obscured and overlooked in the intricacies of Green’s functions \cite{Vorberger2025} and quantum-kinetic \cite{Tolias_2023}  analyses, emerges here with surprising clarity. As a benchmark for future theoretical developments, we provide exact data for $\chi_0^{(1,3)}(\vec{q})$ computed by Kohn-Sham (KS) DFT simulations of the harmonically perturbed ideal uniform electron gas (UEG). Additionally, we derive the exact analytical solution for the long-wavelength limit of $\chi_0^{(1,3)}(\vec{q})$ of the UEG.

We employ the developed framework to evaluate the performance of commonly used non-interacting free-energy functional approximations in describing the non-linear density response of the UEG across temperatures spanning from the ground state to the high-temperature regime. We consider the ground-state Wang-Teter (WT) functional \cite{Wang_Teter} and its finite-temperature version developed by Sjostrom and Daligault \cite{WTF}, which is conventionally referred to as the WTF functional. Additionally, we examine the generalized gradient approximation (GGA) introduced by Luo, Karasiev, and Trickey, denoted as LKTF \cite{LKTF, LKT_T0}. Lastly, we discuss the recently proposed fully nonlocal functional by Ma et al.~\cite{Cheng_PRB_2024, XWM_T0}, referred to as XWMF. The discussion of the results primarily focuses on warm dense matter (WDM) applications, a state of matter characterized by the simultaneous importance of strong thermal excitations, quantum degeneracy effects, and interparticle correlations \cite{Kraus_NatRevPhys_2025, review, wdm_book, Bonitz_pop_2024, Dornheim_review, Moldabekov_PPNP_2025, new_POP}; see Ref.~\cite{vorberger2025roadmapwarmdensematter} for a topical WDM roadmap article.

Because orbital-free methods rely heavily on the UEG limit, the accuracy of the non-interacting free-energy functional in this regime is critical for modeling metals, semiconductors, and quantum plasmas \cite{Wenhui_Chemical_Reviews, Cheng_Comp_Mol_Sci, Moldabekov_prb_2023, Moldabekov_Electronic_Structure_2025, LKTF, Moldabekov_PPNP_2025}. 
The presented analysis can be particularly valuable for the development of new non-interacting free-energy functionals at conditions where accurate modeling of the electronic screening of ion-ion pair interaction is crucial. In this regard, we highlight the previously overlooked fact that the non-interacting free energy functional, which can reproduce the known limit of the ideal linear density response function of the UEG—specifically the Lindhard function \cite{Lindhard1954} (e.g., Wang-Teter-type functionals \cite{Wang_Teter, WTF})—should also automatically reproduce the established analytical solution for the quadratic density response function at the second harmonic $\chi_0^{(2)}(\vec{q})$  \cite{mikhailov2012}. This is because $\chi_0^{(2)}(\vec{q})$ is fully defined in terms of the ideal linear density response function. The same principle applies to higher-order response functions at higher harmonics (the $l$-order ideal density response at the $l$th harmonic \cite{Tolias_2023}), although these become progressively less significant as $l$ increases.

Finally, our results complement a series of earlier works on the non-linear density response of the UEG and related systems~\cite{Mikhailov_Annalen,Mikhailov_PRL,mikhailov2012,Dornheim_PRL_2020,Dornheim_PRR_2021,Dornheim_CPP_2021,Dornheim_CPP_2022,Dornheim_JPSJ_2021,Moldabekov_jctc_2022,Tolias_2023,Vorberger2025,Dornheim_JCP_ITCF_2021,Dornheim_review}, which is important in its own right.

The paper is organized as follows: In Sec.~\ref{s:theory}, we present the theory of non-linear response within the framework of DFT. Sec.~\ref{s:appl} applies this developed theory to analyze non-interacting free energy functionals and examines the behavior of the exact solutions for the ideal non-linear density response functions of the UEG across different parameters. We conclude the paper by summarizing the main findings and providing an outlook on future applications of the presented framework.\\

\section{Theory}\label{s:theory}
\subsection{General Framework}
We start from the free-energy functional of electrons as it is considered in DFT:
\begin{equation}
F[n] = F_s[n] + F_H[n] + F_{\text{xc}}[n]
+ \int d\vec  r\, v_{\text{ext}}(\vec  r)\,n(\vec  r),
\end{equation}
where $F_s[n]$ is the non-interacting free energy functional, $F_H[n]$ is the contribution due to Hartree mean field, $F_{\text{xc}}[n]$ is the contribution due to exchange-correlation effects, and $v_{\text{ext}}$ is an external potential, e.g., due to ions, nuclei and externally applied fields.

For the equilibrium state, we have:
\begin{equation}\label{eq:Fmin}
\frac{\delta F_s}{\delta n(\vec  r)}
+ \frac{\delta F_H}{\delta n(\vec  r)}
+ \frac{\delta F_{\text{xc}}}{\delta n(\vec  r)}
+ v_{\text{ext}}(\vec  r)
= \mu,
\end{equation}
with $\mu$ being the chemical potential playing the role of the Lagrangian multiplier that enforces the particle number conservation \cite{Wenhui_Chemical_Reviews}.

In Eq.~(\ref{eq:Fmin}), it is convenient to define the potentials:
\begin{equation}
v_s(\vec  r) \equiv \frac{\delta F_s}{\delta n(\vec  r)},\quad
v_H(\vec  r) \equiv \frac{\delta F_H}{\delta n(\vec  r)},\quad
v_{\text{xc}}(\vec  r) \equiv \frac{\delta F_{\text{xc}}}{\delta n(\vec  r)}.
\end{equation}

Now, we perturb the initial equilibrium state by applying an external perturbation and consider the change in density $\Delta n$, potentials ($\Delta v_s$ etc.), and chemical potential $\Delta \mu$:
\begin{equation}
\Delta v_s(\vec  r)
+ \Delta v_H(\vec  r)
+ \Delta v_{\text{xc}}(\vec  r)
+ \Delta v_{\text{ext}}(\vec  r)
= \Delta \mu.
\label{eq:realspace_euler}
\end{equation}

We consider a periodic system of volume $\Omega=L^3$ with reciprocal lattice vectors $\mathbf k$. 
The discrete Fourier components are
\begin{equation}
\Delta v(\mathbf r)
= \frac{1}{\Omega} \sum_{\mathbf k} e^{i\mathbf k\cdot\mathbf r}\, \Delta v(\mathbf k),\quad \Delta v(\mathbf k)
= \int_\Omega d\mathbf r\, e^{-i\mathbf k\cdot\mathbf r}\, \Delta v(\mathbf r),
\end{equation}
and
\begin{equation}
\int_\Omega d\mathbf r\, e^{i(\mathbf k - \mathbf k')\cdot\mathbf r}
= \Omega \,\delta_{\mathbf k,\mathbf k'},
\label{eq:orthog}
\end{equation}
where $\delta_{\mathbf k,\mathbf k'}$ is the Kronecker delta on the discrete reciprocal lattice.

 With these conventions, the Fourier transform of Eq.~\eqref{eq:realspace_euler}  gives:
\begin{equation}
\Delta v_s(\vec  k)
+ \Delta v_H(\vec  k)
+ \delta v_{\text{xc}}(\vec  k)
+ \Delta v_{\text{ext}}(\vec  k)
= \Delta \mu \, \Omega\,\delta_{\vec k, 0},
\label{eq:fourier_euler_basic}
\end{equation}
where $\Delta \mu \, \delta_{\vec k, 0}$  only matters in the case $\vec k=0$, and for $\vec k\neq 0$ we set  the r.h.s. of Eq. (\ref{eq:fourier_euler_basic}) to zero.
The particle number conservation $\int \rm{d}\vec  r\,\Delta n(\vec r)=0$ is enforced by tuning $\Delta \mu$, which is a number representing the change in the chemical potential.

We perform a functional Taylor expansion of each potential in powers of $\Delta n$. For example, for $\Delta v_s(\vec  r)$ we write:
\begin{widetext}
\begin{align}
\Delta v_s(\vec  r)
= \int \rm{d}\vec  r_1\, {\mathcal{K}}_s^{(2)}(\vec  r,\vec  r_1)\,\Delta n(\vec  r_1)
&+ \iint {\rm d} \vec  r_1 {\rm d}\vec  r_2\, {\mathcal{K}}_s^{(3)}(\vec  r,\vec  r_1,\vec  r_2)\,
\Delta n(\vec  r_1)\,\Delta n(\vec  r_2)
\nonumber \\
&+ \iiint {\rm d}\vec  r_1 {\rm d}\vec  r_2 {\rm d}\vec  r_3\,
{\mathcal{K}}_s^{(4)}(\vec  r,\vec  r_1,\vec  r_2,\vec  r_3)\,
\Delta n(\vec  r_1)\,\Delta n(\vec  r_2)\,\Delta n(\vec  r_3)
+ \cdots \label{eq:vsr_exp}
\end{align}.
\end{widetext}
where the $l$-th order functional derivative of $F_s[n]$ around the equilibrium density $n_0$ is denoted as:
\begin{align}
{\mathcal{K}}_s^{(l)}(\vec r, \vec r_{1},...,\vec r_{l-1})
&\,\underset{l\geq2}{\mathrel{=\!=}} \,
\left.\frac{\delta^{(l)} F_s}{\delta n(\vec  r)\,\delta n(\vec  r_1)...\delta n(\vec  r_{l-1})}\right|_{n_0}.
\end{align}

Similarly, we define the functional derivatives of the XC functional:
\begin{align}
{\mathcal{K}}_{\rm xc}^{(l)}(\vec r, \vec r_{1},...,\vec r_{l-1})
&\,\underset{l\geq2}{\mathrel{=\!=}} \, 
\left.\frac{\delta^{(l)} F_s}{\delta n(\vec  r)\,\delta n(\vec  r_1)...\delta n(\vec  r_{l-1})}\right|_{n_0}.
\end{align}

Further, for an $l$-th order  kernel $\mathcal{K}^{(l)}(\vec r,\vec r_1,\ldots,\vec r_{l-1})$
defined in a periodic cell of volume $\Omega$,
we associate reciprocal vectors $\vec k, \vec k_1,\ldots,\vec k_{l-1}$ to the
coordinates $\vec r,\vec r_1,\ldots,\vec r_{l-1}$, respectively, and define the Fourier transform:
\begin{align}
&\mathcal{K}^{(l)}(\vec k,\vec k_1,\ldots,\vec k_{l-1})
= \nonumber \\
&\int_{\Omega} d\vec r
\int_{\Omega} d\vec r_1
\cdots
\int_{\Omega} d\vec r_{l-1} \;
e^{-i \vec k \cdot \vec r}
\prod_{j=1}^{l-1}
\left[
e^{+i \vec k_j \cdot \vec r_j}
\right]
\mathcal{K}(\vec r,\vec r_1,\ldots,\vec r_{l-1}).
\end{align}

Next, we explicitly assume a homogeneous system as it is realized in disordered materials \cite{JCP_averaging, Moldabekov_PPNP_2025} (i.e., no crystal lattice). Due to translational invariance, one can introduce reduced kernels:
\begin{equation}\label{eq:Kl_gen}
    \mathcal{K}^{(l)}(\vec k,\vec k_1,\ldots,\vec k_{l-1})=\Omega \delta_{\vec k, \,\vec k_1+\cdots+\vec k_{l-1}}\widetilde{\mathcal{K}}^{(l)}(\vec k_1,\ldots,\vec k_{l-1}).
\end{equation}

 
In Eq.~(\ref{eq:Kl_gen}), let us refer to $\vec k$  as  the  ``output'' wavevector, and to $\left\{\vec k_1,..,\vec k_l\right\}$ as the ``input'' wavevectors.
Furthermore, for bookkeeping purposes, we explicitly indicate in the reduced kernels 
the ``output'' wavevector by introducing the following notation for the kernels with $l>2$:
\vspace{0.25cm}
\begin{equation}
    \widetilde{\mathcal{K}}^{(l)}(\vec k_1,\ldots,\vec k_{l-1})=\widetilde{\mathcal{K}}^{(l)}(\vec k|\,\vec k_1,\ldots,\vec k_{l-1}),
\end{equation}
which simply means that the ``input'' wavevectors in the reduced kernels satisfy the momentum conservation $\vec k=\vec k_1+...+\vec k_{l-1}$.  This notation allows one to group density response terms by harmonics (in accordance with Eq.~(\ref{eq:n_start})) in a transparent way. 

With these notations, a Fourier transform $\Delta v^{(l)}(\vec k)$  of the $l$-th order term $\Delta v^{(l)}(\vec r)$, in the expansions such as Eq.~(\ref{eq:vsr_exp}), can now be expressed as
\begin{widetext}
\begin{align}
\Delta v^{(l)}(\vec k)&\underset{l\geq2}{\mathrel{=\!=}} 
\frac{1}{\Omega^{\,l-2}}
\sum\limits_{\substack{\vec k_1,\ldots,\vec k_{l-1} \\ \vec k=\vec k_1+\ldots\vec k_{l-1}}}
\widetilde{\mathcal{K}}^{(l)}(\vec k|\,\vec k_1,\ldots,\vec k_{l-1})\Delta n(\vec k_1)\cdots\Delta n(\vec k_{l-1}),\label{eq:dvlk_gen}
\end{align}
\end{widetext}
 where we set $l\geq2$  to align with the order of the functional derivatives in the kernels of the expansion, and used the notation
$\sum\limits_{\substack{\vec k_1,\ldots,\vec k_{l-1} \\ \vec k=\vec k_1+\ldots\vec k_{l-1}}}$ meaning that the sum is taken only over those sets of momenta 
$\vec k_1,\ldots,\vec k_{l-1}$ whose total momentum equals the ``input'' 
momentum $\vec k$. This enforces the momentum-conserving condition 
$\vec k = \vec k_1 + \cdots + \vec k_{l-1}$.

Fourier transforming and using Eq.~\eqref{eq:dvlk_gen}, for the functional Taylor expansions of the potential  $\Delta v_s$ and  $\Delta v_{\rm xc}$, we can write:

\begin{align}
&\Delta v_s(\vec  k)
=\widetilde{\mathcal{K}}_s^{(2)}(\vec  k)\Delta n(\vec  k)\nonumber \\
&+\frac{1}{\Omega}
\sum\limits_{\substack{\vec k_1,\vec k_2 \\ \vec k=\vec k_1+\vec k_2}}
\widetilde{\mathcal{K}}_s^{(3)}(\vec  k|\, \vec  k_1,\vec  k_2)
\Delta n(\vec  k_1)\Delta n(\vec  k_2) \nonumber \\
&+\frac{1}{\Omega^2}
\sum\limits_{\substack{\vec k_1,\vec k_2, \vec k_3 \\ \vec k=\vec k_1+\vec k_2+\vec k_3}}
\widetilde{\mathcal{K}}_s^{(4)}(\vec  k|\,\vec  k_1,\vec  k_2,\vec  k_3)
\Delta n(\vec  k_1)\Delta n(\vec  k_2)\Delta n(\vec  k_3)\nonumber \\
&
+ \cdots,
\label{eq:vs_expand}
\end{align}
and 
\begin{align}
&\Delta v_{\rm xc}(\vec  k)
=\widetilde{\mathcal{K}}_{\rm xc}^{(2)}(\vec  k)\Delta n(\vec  k)\nonumber \\
&+\frac{1}{\Omega}
\sum\limits_{\substack{\vec k_1,\vec k_2 \\ \vec k=\vec k_1+\vec k_2}}
\widetilde{\mathcal{K}}_{\rm xc}^{(3)}(\vec  k|\, \vec  k_1,\vec  k_2)
\Delta n(\vec  k_1)\Delta n(\vec  k_2) \nonumber \\
&+\frac{1}{\Omega^2}
\sum\limits_{\substack{\vec k_1,\vec k_2, \vec k_3 \\ \vec k=\vec k_1+\vec k_2+\vec k_3}}
\widetilde{\mathcal{K}}_{\rm xc}^{(4)}(\vec  k|\,\vec  k_1,\vec  k_2,\vec  k_3)
\Delta n(\vec  k_1)\Delta n(\vec  k_2)\Delta n(\vec  k_3)\nonumber \\
&
+ \cdots,
\label{eq:vxc_expand}
\end{align}

Note that $\widetilde{\mathcal{K}}_{\rm xc}^{(2)}(\vec k)$ in Eq.~(\ref{eq:vxc_expand}) is the usual static XC kernel used in linear response time-dependent DFT \cite{book_Ullrich, Moldabekov_PRR_2023, Moldabekov_JCTC_2023}.

For the mean-field Hartree potential, we have:
\begin{equation}
\Delta v_H(\vec  k)
= v_c(\vec  k)\,\Delta n(\vec  k),
\qquad
v_c(\vec  k)=\frac{4\pi}{k^2}.
\label{eq:hartree_linear}
\end{equation}

It is convenient to define the \emph{total} kernels:
\begin{align}\label{eq:k2tot_def}
\widetilde{\mathcal{K}}_{\rm tot}^{(2)}(\vec  k)
&\equiv
\widetilde{\mathcal{K}}_s^{(2)}(\vec  k)
+ v_c(\vec  k)
+  \widetilde{\mathcal{K}}_{\rm xc}^{(2)}(\vec  k),\\
\widetilde{\mathcal{K}}_{\rm tot}^{(3)}(\vec  k|\,\vec  k_1,\vec  k_2)
&\equiv
\widetilde{\mathcal{K}}_s^{(3)}(\vec  k|\,\vec  k_1,\vec  k_2)
+ \widetilde{\mathcal{K}}_{\rm xc}^{(3)}(\vec  k|\,\vec  k_1,\vec  k_2),\\
\widetilde{\mathcal{K}}_{\rm tot}^{(4)}(\vec  k|\,\vec  k_1,\vec  k_2,\vec  k_3)
&\equiv
\widetilde{\mathcal{K}}_s^{(4)}(\vec  k|\,\vec  k_1,\vec  k_2,\vec  k_3)\\
&\quad + \widetilde{\mathcal{K}}_{\rm xc}^{(4)}(\vec  k|\,\vec  k_1,\vec  k_2,\vec  k_3).
\end{align}

Inserting Eqs.~\eqref{eq:vs_expand},  \eqref{eq:vxc_expand}, and \eqref{eq:hartree_linear} into Eq.~\eqref{eq:fourier_euler_basic}, and using notations for the total kernels, for each nonzero $\vec  k$ we finally get:
\begin{align}
&\widetilde{\mathcal{K}}_{\rm tot}^{(2)}(\vec  k)\Delta n(\vec  k)
+\frac{1}{\Omega}
\sum\limits_{\substack{\vec k_1,\vec k_2 \\ \vec k=\vec k_1+\vec k_2}}
\widetilde{\mathcal{K}}_{\rm tot}^{(3)}(\vec  k|\,\vec  k_1,\vec  k_2)
\Delta n(\vec  k_1)\Delta n(\vec  k_2)
\nonumber \\
&
+\frac{1}{\Omega^2}
\sum\limits_{\substack{\vec k_1,\vec k_2, \vec k_3 \\ \vec k=\vec k_1+\vec k_2+\vec k_3}}
\widetilde{\mathcal{K}}_{\rm tot}^{(4)}(\vec  k|\,\vec  k_1,\vec  k_2,\vec  k_3)
\Delta n(\vec  k_1)\Delta n(\vec  k_2)\Delta n(\vec  k_3)\nonumber \\
&
+\,...\,= -\Delta v_{\text{ext}}(\vec  k).
\label{eq:master_compact}
\end{align}

One can rewrite Eq.~\eqref{eq:master_compact} in a general compact form as
\begin{widetext}
\begin{equation}
  \widetilde{\mathcal{K}}_{\rm tot}^{(2)}(\vec  k)\Delta n(\vec  k)+  \sum\limits_{l=3}^{\infty}\left[{\Omega^{\,2-l}}
\sum\limits_{\substack{\vec k_1,\ldots,\vec k_{l-1} \\ \vec k=\vec k_1+\ldots\vec k_{l-1}}}
\widetilde{\mathcal{K}}_{\rm tot}^{(l)}(\vec k|\,\vec k_1,\ldots,\vec k_{l-1})\Delta n(\vec k_1)\cdots\Delta n(\vec k_{l-1})\right]=-\Delta v_{\text{ext}}(\vec  k),
\label{eq:master_compact2}
\end{equation}
where
\begin{equation}
    \widetilde{\mathcal{K}}_{\rm tot}^{(l)}(\vec k|\,\vec k_1,\ldots,\vec k_{l-1})
\,\underset{l>2}{\mathrel{=\!=}}\, 
\widetilde{\mathcal{K}}_s^{(l)}(\vec k|\,\vec k_1,\ldots,\vec k_{l-1})\,
+  \,\widetilde{\mathcal{K}}_{\rm xc}^{(l)}(\vec k|\,\vec k_1,\ldots,\vec k_{l-1}),
\end{equation}
\end{widetext}
and $\widetilde{\mathcal{K}}_{\rm tot}^{(2)}(\vec  k)$ is defined in Eq.~(\ref{eq:k2tot_def}).

If we denote the left-hand side of Eq.~\eqref{eq:master_compact2} as $\Delta v_{\rm ind}(\vec{k})$, we can express it as $\Delta v_{\rm ind}(\vec{k}) = -\Delta v_{\text{ext}}(\vec{k})$, which means that the total induced potential exactly counteracts the external perturbation. Within linear density-response theory, this corresponds to omitting the sum on the left-hand side of Eq.~\eqref{eq:master_compact2}.

For systems that are homogeneous on average, Eq.~\eqref{eq:master_compact2} can be utilized to analyze the density response of electrons to various types of external potentials, provided that the induced density perturbations can be expressed as a functional Taylor expansion around a homogeneous density value of the system. In the following, we will use Eq.~\eqref{eq:master_compact2} to derive the non-linear density response functions to a static harmonic external perturbation.

\subsection{Task-Specific Formulation}

Considering the harmonic perturbation in the form of
\begin{equation}\label{eq:pert}
\Delta v_{\text{ext}}(\vec  r) = 2A \cos(\vec  q\cdot\vec  r)=A e^{i\vec  q\cdot \vec  r} + A e^{-i\vec  q\cdot \vec  r},
\end{equation}
we now solve Eq.~\eqref{eq:master_compact2}  order by order in $A$. 

The discrete Fourier transform of $\Delta v_{\text{ext}}(\vec  r)$ reads:
\begin{equation}
    \Delta v_{\text{ext}}(\vec  k)=A \Omega\left[\delta_{\vec k,\vec q}+\delta_{\vec k,-\vec q}\right].
\end{equation}

 External perturbation (\ref{eq:pert}) induces a density perturbation represented by a harmonic expansion \cite{Mikhailov_Annalen, Dornheim_PRR_2021}
\begin{align}
\Delta n(\vec  r)
=& 2 \, \rho{(\vec q)} \cos(\vec  q\cdot \vec  r) \nonumber\\ 
&+ 2\,\rho (2\vec q) \cos(2\vec  q \cdot \vec  r)
+ 2\,\rho(3\vec q) \cos(3\vec  q \cdot \vec  r)
+ \cdots,\label{eq:n_start}
\end{align}
with the discrete Fourier transform:
\begin{equation}
    \Delta n(\vec k)=\Omega
    \sum_{m}\rho(m\, \vec q)\left[\delta_{\vec k,\,m\vec q}+\delta_{\vec k,\,-m\vec q}\right],
    \label{eq:dnk_gen}
\end{equation}
where we consider solutions with $\rho(\vec q) = \rho^{(1)}(\vec q)+\rho^{(1,3)}(\vec q)=\mathcal O(A)+\mathcal O(A^3)$, $\rho(2\vec q) = \mathcal O(A^2)$, and $\rho(3\vec q) = \mathcal O(A^3)$. 
Factor two in Eq.~(\ref{eq:pert}) and Eq.~(\ref{eq:n_start}) is a convention \cite{Dornheim_PRL_2020, Dornheim_PRR_2021, Moldabekov_jctc_2022, Moldabekov_PPNP_2025}.

The task is to use the DFT framework to find the density response functions defined as \cite{Mikhailov_PRL, Mikhailov_Annalen, moroni, Dornheim_PRL_2020, Dornheim_PRR_2021, Moldabekov_jctc_2022, JCP_averaging, Tolias_2023}
\begin{equation}\label{eq:def_chi13}
    \chi^{(1)}(\vec q)=\frac{\rho^{(1)}{(\vec q)}}{A}, \quad  \chi^{(1,3)}(\vec q)=\frac{\rho^{(1,3)}{(\vec q)}}{A^3},
\end{equation}
and 
\begin{equation}
    \chi^{(2)}(\vec q)=\frac{\rho{(2\vec q)}}{A^2}, \quad  \chi^{(3)}(\vec q)=\frac{\rho{(3\vec q)}}{A^3},
\end{equation}
where $\chi^{(1)}(\vec q)$ is the linear density response function, $\chi^{(1, 3)}(\vec q)$ is the cubic density response function at the first harmonic, $\chi^{(2)}(\vec q)$ is the quadratic density response function at the second harmonic, and $\chi^{(3)}(\vec q)$ is the cubic density response function at the third harmonic.

\subsection{Linear response at $\vec k=\vec q$}
As an introductory exercise, we first recover known results for the linear density response function. For $\vec  k=\vec  q$, at first order in $A$,  we have $\Delta n(\vec q)\simeq \Delta n^{(1)}(\vec q)=\rho^{(1)}{(\vec q)}\Omega$ (the term with $m=1$ in Eq.~(\ref{eq:dnk_gen})). In this case, following \eqref{eq:master_compact2}, we write:
\begin{equation}
\widetilde{\mathcal{K}}_{\rm tot}^{(2)}(\vec  q)\,\Delta n^{(1)}(\vec q)
= -\,\Delta v_{\text{ext}}(\vec  q),
\end{equation}
from which, taking into account that $\Delta v_{\text{ext}}(\vec  q)
= A \, \Omega$ and $\Delta n^{(1)}(\vec q)=\rho^{(1)}(\vec  q)\, \Omega$, we find a known result for the linear density response function:
\begin{align}\label{eq:chi_def1}
\chi^{(1)}(\vec  q) &\equiv \frac{\delta n^{(1)}(\vec q)}{\delta v_{\rm ext}(\vec q)}
=\frac{\rho^{(1)} (\vec q)}{A}
\\
&= -\frac{1}{\widetilde{\mathcal{K}}_{\rm tot}^{(2)}(\vec  q)}
= -\frac{1}{\widetilde{\mathcal{K}}_s^{(2)}(\vec  q)+v_c(\vec  q)+\widetilde{\mathcal{K}}_{\rm xc}^{(2)}(\vec  q)}.
\label{eq:chi_def}
\end{align}

By symmetry $\delta n^{(1)}(-\vec  q) = \delta n^{(1)}(\vec  q)$ and $\chi^{(1)}(-\vec  q) =\chi^{(1)}(\vec  q) $. 

It is a convention to denote
\begin{equation}\label{eq:lindhard}
    \chi^{(1)}_0(\vec q)=-\frac{1}{\widetilde{\mathcal{K}}_s^{(2)}(\vec  q)},
\end{equation}
with $\chi_{0}(\vec q)$ being referred to as the non-interacting density response. 
This function $\chi_{0}(\vec q)$ is given by the Lindhard function for the UEG and by the KS-response function in DFT \cite{Kollmar, JCP_averaging}. 

Using Eq.~(\ref{eq:lindhard}) in Eq.~(\ref{eq:chi_def}), we arrive at a standard form of the linear density response function from the theory of quantum liquids \cite{quantum_theory}:
\begin{equation}\label{eq:lind}
    \chi^{(1)}(\vec  q) 
= \frac{\chi^{(1)}_0(\vec q)}{1-\left(v_c(\vec  q)+\widetilde{\mathcal{K}}_{\rm xc}^{(2)}(\vec  q)\right)\chi^{(1)}_0(\vec q)}.
\end{equation}

The random phase approximation (RPA) follows from Eq.~(\ref{eq:lind}) if we neglect XC contributions by setting $\widetilde{\mathcal{K}}_{\rm xc}^{(2)}(\vec  q)\equiv0$ (denoted as $ \chi_{\rm RPA}^{(1)}(\vec  q)$).

\subsection{Quadratic response at second harmonic $\vec k= 2\vec q$}

Next, we compute the response at $2\vec  q$, which emerges to be $\mathcal O(A^2)$.
For that, we set $\vec  k=2\vec  q$ in Eq.~\eqref{eq:master_compact2}, where now $\Delta v_{\text{ext}}(2\vec  q)=0$ because there is no direct external perturbation at $2\vec  q$. In Eq.~\eqref{eq:master_compact2}, we  consider terms up to $\mathcal O(A^2)$. Taking into account momentum conservation, we set $\vec k_1 = \vec q$ and $\vec k_2 = \vec q$,  resulting from the condition $2\vec q = \vec k_1 + \vec k_2$. Other combinations would yield higher-order terms in $A$. Additionally, we exclude terms with $\vec k_1 = 0$ or $\vec k_2 = 0$ as they cancel out due to the conservation of the number of particles, $\Delta n(\vec k=0)=\int_\Omega \rm{d}\vec  r\,\Delta n(\vec r)=0$. Therefore, Eq.~\eqref{eq:master_compact2} gives us the quadratic response of the form:
\begin{equation}
\widetilde{\mathcal{K}}_{\rm tot}^{(2)}(2\vec  q)\,\Delta n(2\vec  q)
+\frac{1}{\Omega}
\widetilde{\mathcal{K}}_{\rm tot}^{(3)}(2\vec  q|\, \vec  q,\vec  q)\,
\Delta n(\vec  q)\,\Delta n(\vec  q)
= 0.
\label{eq:k2q}
\end{equation}

Eq.~\eqref{eq:k2q} then leads to:
\begin{equation}\label{eq:dn2}
\Delta n(2\vec  q)
= -\,\frac{\widetilde{\mathcal{K}}_{\rm tot}^{(3)}(2\vec  q|\, \vec  q,\vec  q)}{\widetilde{\mathcal{K}}_{\rm tot}^{(2)}(2\vec  q)}\,\frac{\left(\Delta n(\vec  q)\right)^2}{\Omega}
.
\end{equation}

From Eq.~(\ref{eq:dn2}), by using $\Delta n(\vec q)\simeq \Delta n^{(1)}(\vec q)$, along with Eqs.~(\ref{eq:chi_def1}) and (\ref{eq:chi_def}), and taking into account that $\Delta n(2\vec q)=\Omega \rho({2\vec q})$ and $\Delta v_{\rm ext}(\vec q)=A\Omega$, we derive
\begin{equation}\label{eq:dn2_fin}
\rho({2\vec q})
= \widetilde{\mathcal{K}}_{\rm tot}^{(3)}(2\vec  q|\,\vec  q,\vec  q)
\chi^{(1)}(2\vec  q)\,\left[\chi^{(1)}(\vec  q)\right]^2 A^2,
\end{equation}
from which we find:
\begin{align}
    \chi^{(2)}(\vec q)\equiv \frac{\rho({2\vec q})}{A^2}=\widetilde{\mathcal{K}}_{\rm tot}^{(3)}(2\vec  q|\,\vec  q,\vec  q)
\chi^{(1)}(2\vec  q)\,\left[\chi^{(1)}(\vec  q)\right]^2.\label{eq:chi2_tot}
\end{align}

Solution \eqref{eq:chi2_tot} for the quadratic response is applicable for any disordered material for which averaging over ion configurations (snapshots) leads to a homogeneous electron density. Let us analyze Eq. \eqref{eq:chi2_tot} in the limit of the UEG and show the connection with the results from previous works.

Eq.~(\ref{eq:k2q}) allows us to understand the origin of the quadratic density response at the second harmonic. For that we rewrite Eq.~(\ref{eq:k2q}) as
\begin{equation}
\widetilde{\mathcal{K}}_{\rm tot}^{(2)}(2\vec  q)\,\Delta n(2\vec  q)
= v^{(2)}_{\rm ind}(2\vec q),
\label{eq:k2q_eff}
\end{equation}
where 
\begin{equation}
    v^{(2)}_{\rm ind}(2\vec q)=\frac{1}{\Omega}
\widetilde{\mathcal{K}}_{\rm tot}^{(3)}(2\vec  q|\, \vec  q,\vec  q)\,
\Delta n(\vec  q)\,\Delta n(\vec  q).
\label{eq:Vind2_eff}
\end{equation}

From Eqs.~(\ref{eq:k2q_eff}) and (\ref{eq:Vind2_eff}), we see that the quadratic density response at the second harmonic is the linear reaction to the second-order field $v^{(2)}_{\rm ind}(2\vec q)$ from the expansion of the perturbation of the total induced potential.

Considering the non-interacting UEG with $\widetilde{\mathcal{K}}_{\rm tot}^{(3)}\to\widetilde{\mathcal{K}}_{s}^{(3)}$,  $\chi^{(2)}\to\chi_0^{(2)}$, and $\chi^{(1)}\to\chi_0$, from Eq.~(\ref{eq:chi2_tot}) we derive:
\begin{equation}\label{eq:Ks3_chi0}
    \widetilde{\mathcal{K}}_{\rm s}^{(3)}(2\vec  q|\,\vec  q,\vec  q)=\frac{\chi_0^{(2)}(\vec q)}{\chi^{(1)}_0(2\vec  q)\,\left[\chi^{(1)}_0(\vec  q)\right]^2}\,,
\end{equation}
where  $\chi_0^{(2)}(\vec q)$ was shown to be expressed in terms of the ideal linear density response function $\chi^{(1)}_0(\vec q)$  by Mikhailov \cite{Mikhailov_PRL}:
\begin{equation}\label{eq:chi2_0}
    \chi_0^{(2)}(\vec q)=\frac{2}{q^2}\left(\chi^{(1)}_0(2\vec  q)-\chi^{(1)}_0(\vec q)\right).
\end{equation}

Now,  for $ \chi^{(2)}(\vec q)$ of the UEG, we find:
\begin{widetext}
\begin{equation}
        \chi^{(2)}(\vec q)=\left[\frac{\chi_0^{(2)}(\vec q)}{\chi^{(1)}_0(2\vec  q)\,\left[\chi^{(1)}_0(\vec  q)\right]^2}+\widetilde{\mathcal{K}}_{\rm xc}^{(3)}(2\vec  q|\,\vec  q,\vec  q)\right]
\chi^{(1)}(2\vec  q)\,\left[\chi^{(1)}(\vec  q)\right]^2,\label{eq:chi2_tot2}
\end{equation}
\end{widetext}
from which one can see that full inclusion of the XC effects in $\chi^{(2)}(\vec q)$ of the UEG requires  $\widetilde{\mathcal{K}}_{\rm xc}^{(3)}(2\vec  q|\,\vec  q,\vec  q)$, i.e., the third-order functional derivative of the XC functional \cite{Ashcroft_prl_2003}.  

If we neglect the contribution $\widetilde{\mathcal{K}}_{\rm xc}^{(3)}(2\vec  q|\,\vec  q,\vec  q)$, Eq.~(\ref{eq:chi2_tot2}) reproduces the result derived in Ref.~\cite{Dornheim_PRR_2021}:
\begin{align}\label{eq:chi2_old}
            \chi^{(2)}(\vec q)=&{\chi_0^{(2)}(\vec q)}\left[1-v_c(\vec  q)\left[1-G(\vec  q)\right]\chi^{(1)}_0(\vec q)\right]^{-2}\nonumber \\
            &\times \left[1-v_c(2\vec  q)\left[1-G(2\vec  q)\right]\chi^{(1)}_0(2\vec  q)\right]^{-1},
\end{align}
where we applied Eq.~(\ref{eq:lind}) and introduced the local field correction $G(\vec q) = -\widetilde{\mathcal{K}}_{\rm xc}^{(2)}(\vec q) / v_c(\vec q)$, which is frequently used in the theory of the UEG and quantum liquids~\cite{quantum_theory,kugler1,dornheim_ML,Dornheim_MRE_2024}, instead of the kernel $\widetilde{\mathcal{K}}_{\rm xc}^{(2)}(\vec q)$.  Thus, approximation \eqref{eq:chi2_old} neglects $\widetilde{\mathcal{K}}_{\rm xc}^{(3)}(2\vec  q|\,\vec  q,\vec  q)$. 

Within RPA, i.e., setting $\widetilde{\mathcal{K}}_{\rm xc}^{(3)}(2\vec  q|\,\vec  q,\vec  q)=0$ and $\widetilde{\mathcal{K}}_{\rm xc}^{(2)}(\vec q)=0$, we find from Eq.~(\ref{eq:chi2_tot2}):
\begin{align}\label{eq:chi2_rpa}
            \chi_{\rm RPA}^{(2)}(\vec q)=&{\chi_0^{(2)}(\vec q)}\left[1-v_c(\vec  q)\chi^{(1)}_0(\vec q)\right]^{-2}\nonumber \\
            &\times \left[1-v_c(2\vec  q)\chi^{(1)}_0(2\vec  q)\right]^{-1},
\end{align}
which agrees with the result by  derived using the method of Green’s function (e.g., see Refs.~\cite{Vorberger2025, Pitarke_prb_1995, Zaremba_prb_1988} ).

One can express $\widetilde{\mathcal{K}}_{\rm xc}^{(3)}(2\vec  q|\,\vec  q,\vec  q)$ in terms of the linear and quadratic density response functions:
\begin{equation}
    \widetilde{\mathcal{K}}_{\rm xc}^{(3)}(2\vec  q|\,\vec  q,\vec  q)=\frac{\chi^{(2)}(\vec q)}{\chi^{(1)}(2\vec  q)\,\left[\chi^{(1)}(\vec  q)\right]^2}-\frac{\chi_0^{(2)}(\vec q)}{\chi^{(1)}_0(2\vec  q)\,\left[\chi^{(1)}_0(\vec  q)\right]^2}.\label{eq:kxc3}
\end{equation}

Eq.~(\ref{eq:kxc3}) allows one to compute $\widetilde{\mathcal{K}}_{\rm xc}^{(3)}(2\vec  q|\,\vec q,\vec q)$ without performing third order functional derivatives by using the data for $\chi^{(2)}(\vec q)$, $\chi_0^{(2)}(\vec q)$, $\chi_0^{(2)}(\vec q)$, and $\chi^{(1)}_0(\vec q)$ from the simulation of electrons perturbed by field (\ref{eq:pert}). Such an approach was extensively used to compute the static XC kernel $\widetilde{\mathcal{K}}_{\rm xc}^{(2)}(\vec q)$ for the UEG and warm dense hydrogen to analyze various XC functionals often used DFT \cite{Moldabekov_JCTC_2023, Moldabekov_PRR_2023, Moldabekov_non_empirical_hybrid, Moldabekov_PRB_2022, Moldabekov_jcp_2023_hyb, Moldabekov_JCP_2021,Bohme_PRL_2022}. 
Similarly, Eq.~(\ref{eq:kxc3}) can be used to analyze XC functionals on the level of the third-order functional derivatives. 
For example, exact quantum Monte Carlo data for  $\widetilde{\mathcal{K}}_{\rm xc}^{(3)}(2\vec  q|\,\vec  q,\vec  q)$ of the UEG can be used as an additional constraint for the development of new XC functionals with improved consistency.

\subsection{Cubic response at the first harmonic $\vec k=\vec q$}

Next, we examine the cubic response at the first harmonic, denoted as $\chi^{(1,3)}(\vec q)$ and  defined in Eq. \eqref{eq:def_chi13}. Previous attempts to find an analytic solution for $\chi^{(1,3)}(\vec q)$, utilizing the method of Green’s functions \cite{Vorberger2025} and quantum-kinetic theory \cite{Tolias_2023}, were unsuccessful, leading to a conspicuous absence of theoretical results for this case.
Here, we conclusively overcome this problem using the present DFT framework.

First, in addition to the linear response term,  we include the  $\mathcal O(A^3)$ term in $\Delta n(\vec k=\vec  q)$:
\begin{equation}
\Delta n(\vec  q)
= \Delta n^{(1)}(\vec  q) + \Delta n^{(3)}(\vec  q),
\end{equation}
where $\Delta n^{(1)}(\vec  q)=\chi^{(1)}(\vec q)\, A \,\Omega$,  and $\Delta n^{(3)}(\vec  q)=\chi^{(1,3)}(\vec q)\, A^3\, \Omega$ is to be found.

We set $\vec  k=\vec  q$ in Eq.~\eqref{eq:master_compact2}, and expand to $\mathcal O(A^3)$. The left-hand side has three pieces:
\begin{enumerate}
\item  For the linear term, we have:
\begin{equation}
\widetilde{\mathcal{K}}_{\rm tot}^{(2)}(\vec  q)\Delta n(\vec  q)= \widetilde{\mathcal{K}}_{\rm tot}^{(2)}(\vec  q)
\left[
\Delta n^{(1)}(\vec  q)+\Delta n^{(3)}(\vec  q)
\right],
\end{equation}
which, at $\mathcal O(A)$, matches $-\Delta v_{\text{ext}}(\vec  q)$ on the r.h.s. of Eq.~\eqref{eq:master_compact2}, and has already been  considered. Therefore, only the piece
$\widetilde{\mathcal{K}}_{\rm tot}^{(2)}(\vec  q)\,\Delta n^{(3)}(\vec  q)\sim\mathcal O(A^3)$ remains.

\item In the quadratic term
\begin{equation}
\quad \frac{1}{\Omega}
\sum\limits_{\substack{\vec k_1,\vec k_2 \\ \vec q=\vec k_1+\vec k_2}}
\widetilde{\mathcal{K}}_{\rm tot}^{(3)}(\vec  q|\, \vec  k_1,\vec  k_2)\,
\Delta n(\vec  k_1)\,\Delta n(\vec  k_2),
\end{equation}
to get $\mathcal O(A^3)$, we must use one mode of order $A^2$ and one of order $A$.
The only such combination of $(\vec k_1, \vec k_2)$ that sums to $\vec  q$ at this order is
$(2\vec  q,-\vec  q)$ and $(-\vec  q,2\vec  q)$.
Both give the same contribution by symmetry,  yielding
\begin{equation}
 \frac{2}{\Omega}\, \widetilde{\mathcal{K}}_{\rm tot}^{(3)}(\vec  q|\, 2\vec  q,-\vec  q)\,
\Delta n(2\vec  q)\,\Delta n(-\vec  q).
\end{equation}

\item In the cubic term, we consider
\begin{equation}
\frac{1}{\Omega^2}
\sum\limits_{\substack{\vec k_1,\vec k_2, \vec k_3 \\ \vec q=\vec k_1+\vec k_2+\vec k_3}}
\widetilde{\mathcal{K}}_{\rm tot}^{(4)}(\vec  q|\,\vec  k_1,\vec  k_2,\vec  k_3)\,
\Delta n(\vec  k_1)\,\Delta n(\vec  k_2)\,\Delta n(\vec  k_3).
\end{equation}
with  three fundamental modes
$\Delta n \propto  A$. The only triples $(\vec k_1, \vec k_2, \vec k_3)$ that sum to $\vec  q$
are permutations of $(\vec  q,\vec  q,-\vec  q)$.
There are three such permutations. Thus, this cubic part becomes
\begin{equation}
\frac{3}{\Omega^2}\, 
\widetilde{\mathcal{K}}_{\rm tot}^{(4)}(\vec  q|\,\vec  q,\vec  q,-\vec  q)\,
\Delta n(\vec  q)\,\Delta n(\vec  q)\,\Delta n(-\vec  q).
\end{equation}
\end{enumerate}

Putting all of this together, the $\vec  k=\vec  q$ component  of Eq.~\eqref{eq:master_compact2}  at $\mathcal O(A^3)$ becomes
\begin{align}
&\widetilde{\mathcal{K}}_{\rm tot}^{(2)}(\vec  q)\,\Delta n^{(3)}(\vec  q)
+ \frac{2}{\Omega}\widetilde{\mathcal{K}}_{\rm tot}^{(3)}(\vec  q|\,2\vec  q,-\vec  q)\,
\Delta n(2\vec  q)\,\Delta n(-\vec  q)\nonumber\\
&+ \frac{3}{\Omega^2}\,
\widetilde{\mathcal{K}}_{\rm tot}^{(4)}(\vec  q|\,\vec  q,\vec  q,-\vec  q)\,
\left[\Delta n(\vec  q)\right]^2\,\Delta n(-\vec  q)
=0.
\label{eq:qA3balance}
\end{align}

Solving Eq.~\eqref{eq:qA3balance} for $\Delta n^{(3)}(\vec  q)$, one derives:
\begin{align}
\Delta n^{(3)}(\vec  q)
= &-\,\frac{1}{\widetilde{\mathcal{K}}_{\rm tot}^{(2)}(\vec  q)}
\Big[
\frac{2}{\Omega}\widetilde{\mathcal{K}}_{\rm tot}^{(3)}(\vec  q|\,2\vec  q,-\vec  q)\,
\Delta n(2\vec  q)\,\Delta n(-\vec  q)\nonumber\\
&+ \frac{3}{\Omega^2}\,
\widetilde{\mathcal{K}}_{\rm tot}^{(4)}(\vec  q|\,\vec  q,\vec  q,-\vec  q)\,
\left[\Delta n(\vec  q)\right]^2\,\Delta n(-\vec  q)
\Big],
\label{eq:deltan3q_raw}
\end{align}
where we substitute $\Delta n(\vec  q)= \chi^{(1)}(\vec  q)\, A\Omega$, $\Delta n(-\vec  q)= \chi^{(1)}(\vec  q)A\Omega$, $\Delta n(2\vec q)=\Omega \rho({2\vec q})=\Omega \chi^{(2)}(\vec q)A^2$, and Eq.~(\ref{eq:dn2_fin}), to write for the first product:
\begin{align}
&\Delta n(2\vec  q)\,\Delta n(-\vec  q)
= \rho({2\vec q}) \chi^{(1)}(\vec  q)A\Omega^2 \nonumber\\
&=\chi^{(2)}(\vec q)\chi^{(1)}(\vec  q)A^{3}\Omega^2,  
\\
&=\widetilde{\mathcal{K}}_{\rm tot}^{(3)}(2\vec  q|\,\vec  q,\vec  q)\,
\chi^{(1)}(2\vec  q)\,[\chi^{(1)}(\vec  q)]^{3}\,A^{3}\Omega^2,
\label{eq:prod1}
\end{align}
and for the second product:
\begin{align}
\left[\Delta n(\vec  q)\right]^2\,\Delta n(-\vec  q)
&=\left[\chi^{(1)}(\vec  q)\right]^3 A^3\Omega^3.
\label{eq:prod2}
\end{align}

Substituting Eqs.~\eqref{eq:prod1} and \eqref{eq:prod2} into Eq.~\eqref{eq:deltan3q_raw},  and taking into account that $\chi^{(1)}(\vec  q)=-1/\widetilde{\mathcal{K}}_{\rm tot}^{(2)}(\vec  q)$, we obtain:
\begin{align}
\Delta n^{(3)}(\vec  q)
=
& A^3 \Omega\left[\chi^{(1)}(\vec  q)\right]^4
\left[
3\,\widetilde{\mathcal{K}}_{\rm tot}^{(4)}(\vec  q|\,\vec  q,\vec  q,-\vec  q)\right.\nonumber\\
&\left. +2\,\widetilde{\mathcal{K}}_{\rm tot}^{(3)}(\vec  q|\,2\vec  q,-\vec  q)\,
\widetilde{\mathcal{K}}_{\rm tot}^{(3)}(2\vec  q|\,\vec  q,\vec  q)\,
\chi^{(1)}(2\vec  q)
\right],
\label{eq:deltan3q_final}
\end{align}
from which, using $\Delta n^{(3)}(\vec  q)=\rho^{(1,3)}(\vec q)\,\Omega$, we finally get the cubic response at the first harmonic:
\begin{align}\label{eq:chi13_new}
    \chi^{(1,3)}(\vec q) \equiv& \frac{\rho^{(1,3)}(\vec q)}{A^3} \nonumber \\
    =&\,\left[\chi^{(1)}(\vec  q)\right]^4\Bigg[
3\,\widetilde{\mathcal{K}}_{\rm tot}^{(4)}(\vec  q|\,\vec  q,\vec  q,-\vec  q) \nonumber\\
&\ +2\,\widetilde{\mathcal{K}}_{\rm tot}^{(3)}(\vec  q|\,2\vec  q,-\vec  q)\,
\widetilde{\mathcal{K}}_{\rm tot}^{(3)}(2\vec  q|\,\vec  q,\vec  q)\,
\chi^{(1)}(2\vec  q)\Bigg],
\end{align}
which is our general solution for $\chi^{(1,3)}(\vec q)$. 

To understand the origin of the cubic response at first harmonic, one can rewrite Eq.~(\ref{eq:chi13_new}) as
\begin{align}\label{eq:chi13_new2}
    \rho^{(1,3)}(\vec q)
    =&\chi^{(1)}(\vec  q)\left(\left[A_{\rm ind}(\vec q, \vec q,-\vec q)\right]^3+\left[A_{\rm ind}(2\vec q, -\vec q)\right]^3\right)
\end{align}
where
\begin{align}\label{eq:chi13_magnitude}
\left[A_{\rm ind}(\vec q, \vec q,-\vec q)\right]^3
    =&
3\widetilde{\mathcal{K}}_{\rm tot}^{(4)}(\vec  q|\vec  q,\vec  q,-\vec  q)\left[\chi^{(1)}(\vec  q)\right]^2\chi^{(1)}(-\vec  q)A^3.
\end{align}
and
\begin{align}\label{eq:chi13_magnitude2}
  \left[A_{\rm ind}(2\vec q, -\vec q)\right]^3
    =&2\,\widetilde{\mathcal{K}}_{\rm tot}^{(3)}(\vec  q|\,2\vec  q,-\vec  q)\,
\chi^{(2)}(\vec q)\chi^{(1)}(-\vec  q)\,A^3.
\end{align}

Physically, Eq.~(\ref{eq:chi13_magnitude}) tells us that the external perturbation at wavevector $\vec k=\vec q$ induces three $\mathcal O(A)$-order density perturbations at modes $(\vec q, \vec q, -\vec q)$  that couple with each other to generate a $\mathcal O(A^3)$-order response at the first harmonic (with the coupling being defined by the kernel $\widetilde{\mathcal{K}}_{\rm tot}^{(4)}(\vec  q|\,\vec  q,\vec  q,-\vec  q)$). Eq.~(\ref{eq:chi13_magnitude2}) tells us that the external perturbation at wavevector $\vec k=\vec q$ also induces a $\mathcal O(A^2)$-order density perturbation at $2\vec q$  that couples with the $\mathcal O(A)$-order density perturbation at $-\vec q$ (with the coupling being given by the kernel $\widetilde{\mathcal{K}}_{\rm tot}^{(3)}(\vec q|\,2\vec  q,-\vec q)$) that generates the $\mathcal O(A^3)$ order response at the first harmonic.
Thus, density perturbation  (\ref{eq:chi13_new2}) is the result of the mode coupling of fields generated by induced densities at the second and first harmonics. 


Our new result (\ref{eq:chi13_new}) provides a solution in terms of the functional derivatives of $F_s[n]$ and $F_{\rm xc}[n]$. Although the exact forms of both $F_s[n]$ and $F_{\rm xc}[n]$ are unknown, there are a number of well-tested approximations for $F_s[n]$ and $F_{\rm xc}[n]$ (e.g., for finite temperatures, see Refs.~\cite{groth_prl, Karasiev_prl_2014} and, for the ground state, see Refs.~\cite{Perdew_Wang, Perdew_Zunger_PRB_1981}). In addition, the non-interacting free energy is computed exactly in KS-DFT, allowing for the computation of the density response functions using the direct perturbation approach without relying on approximations for $F_s[n]$ \cite{Moldabekov_jctc_2022, Moldabekov_JCTC_2023}. 
Furthermore, for the UEG, at small wavenumbers $\vec q \to 0$, the exact solution for $F_s[n]$ is given by {the Thomas-Fermi solution for the free energy} $F_{\rm TF}[n]$~\cite{Perrot, Kirzhnits_1975}.

Let us now analyze $\chi^{(1,3)}(\vec q)$ in the limit of the UEG.
For that, we first denote
\begin{align}
    \Gamma_{\rm tot}^{(1,3)}(\vec q)&=3\,\widetilde{\mathcal{K}}_{\rm tot}^{(4)}(\vec  q|\,\vec  q,\vec  q,-\vec  q)\nonumber\\
    &+2\,\widetilde{\mathcal{K}}_{\rm tot}^{(3)}(\vec  q|\,2\vec  q,-\vec  q)\,
\widetilde{\mathcal{K}}_{\rm tot}^{(3)}(2\vec  q|\,\vec  q,\vec  q)\,
\chi^{(1)}(2\vec  q).
\end{align}

In the case of the ideal UEG,
neglecting XC terms, we have
\begin{align}
    \Gamma_{s}^{(1,3)}(\vec q)&=3\,\widetilde{\mathcal{K}}_{s}^{(4)}(\vec  q|\,\vec  q,\vec  q,-\vec  q)\nonumber\\
    &+2\,\widetilde{\mathcal{K}}_{s}^{(3)}(\vec  q|\,2\vec  q,-\vec  q)\,
\widetilde{\mathcal{K}}_{s}^{(3)}(2\vec  q|\,\vec  q,\vec  q)\,
\chi_0^{(1)}(2\vec  q).
\end{align}

The exact $\vec q\to 0$ limit of $F_s[n]$ is given by the TF non-interacting free energy functional $F_{\rm TF}[n]$, for which the Fourier transforms of the functional derivatives do not depend on $\vec q$. In addition, we recall that the  $\vec q\to 0$ limit of $\chi_0^{(1)}(\vec q)$ is a finite number.  Thus, we can write:
\begin{equation}
    \Gamma_{s}^{(1,3)}(\vec q\to 0)=\Gamma_{\rm TF}^{(1,3)},
\end{equation}
and find
\begin{equation}
     \chi_0^{(1,3)}(\vec q\to 0)=\left[\chi_0^{(1)}(\vec q\to  0)\right]^4 \times \Gamma_{\rm TF}^{(1,3)}.\label{eq:chi0130}
\end{equation}

The reduced kernels of the TF model for calculating the long wavelength limit of the considered non-linear response functions are provided in Appendix~\ref{s:TF_kernels}.

As an approximation for $\chi_0^{(1,3)}(\vec q)$, one can use the TF approximation for $\widetilde{\mathcal{K}}_{s}^{(3)}$ in Eq.~(\ref{eq:chi13_new}), and find:
\begin{align}\label{eq:chi13_app}
    \chi_0^{(1,3)}(\vec q)
    \approx &\left[\chi_0^{(1)}(\vec  q)\right]^4\Bigg[
3\,\widetilde{\mathcal{K}}_{\rm TF}^{(4)} \ +2\,\left[\widetilde{\mathcal{K}}_{\rm TF}^{(3)}\right]^2\,
\chi_0^{(1)}(2\vec  q)\Bigg].
\end{align}

This approximation can be expected to give an adequate $q$ dependence of $\chi_0^{(1,3)}(\vec q)$ at small wavenumbers.
We test the quality of this approximation by comparing it with the results of exact numerical calculations for the UEG.


\subsection{Cubic response at the third harmonic $\vec  k = 3\vec q$}

Lastly, we consider the $\mathcal O(A^3)$ terms at \(\vec k=3\vec q\) in  Eq.~\eqref{eq:master_compact2}:
\begin{align}
K_2^{\text{tot}}(3\vec q)\,\Delta n(3\vec q)
&+ \frac{1}{\Omega}\widetilde{\mathcal{K}}_{\rm tot}^{(3)}(3\vec q|\,2\vec q,\vec q)\,
\Delta n(2\vec q)\,\Delta n(\vec q) \nonumber \\
&+\frac{1}{\Omega^2} \widetilde{\mathcal{K}}_{\rm tot}^{(4)}(3\vec q|\,\vec q,\vec q,\vec q)\,
\Delta n(\vec q)^3
= 0,
\label{eq:chi_cub_st}
\end{align}
where we took into account that  $\Delta v_{\rm ext}(3\vec  q)=0$,  a $\widetilde{\mathcal{K}}_{\rm tot}^{(3)}$ term mixing $\Delta n(2\vec  q)$ (order $A^2$) and $\Delta n(\vec  q)$ (order $A$), because $2\vec  q+\vec  q=3\vec  q$,
and a $\widetilde{\mathcal{K}}_{\rm tot}^{(4)}$ term mixing $\Delta n(\vec  q)\Delta n(\vec  q)\Delta n(\vec  q)$, because $\vec  q+\vec  q+\vec  q=3\vec  q$.

Solving Eq.~(\ref{eq:chi_cub_st}) for $\Delta n(3\vec q)$, one finds:
\begin{align}
    \Delta n(3\vec q)
=& -\,\frac{1}{K_2^{\text{tot}}(3\vec q)}\left[
 \frac{1}{\Omega}\widetilde{\mathcal{K}}_{\rm tot}^{(3)}(3\vec q|\,2\vec q,\vec q)\,
\Delta n(2\vec q)\,\Delta n(\vec q)\right. \nonumber \\
&\left.+  \frac{1}{\Omega^2}\widetilde{\mathcal{K}}_{\rm tot}^{(4)}(3\vec q|\,\vec q,\vec q,\vec q)\,
\left[\Delta n(\vec q)\right]^3
\right],\label{eq:chi_cub_inter}
\end{align}
which shows the contribution originated from the coupling of three $\mathcal O(A)$-order density perturbations at modes $(\vec q, \vec q, \vec q)$ and another contribution from the coupling of  $\mathcal O(A^2)$-order density perturbation at $2\vec q$ with $\mathcal O(A)$-order density perturbation at $\vec q$. 

Substituting $\Delta n(\vec  q)\simeq \Delta n^{(1)}(\vec  q)= \chi^{(1)}(\vec  q) A\Omega$, $\Delta n(2\vec  q)=\Omega \rho(2\vec q)$, into Eq.~(\ref{eq:chi_cub_inter}),  taking into account Eq.~(\ref{eq:dn2_fin}) and that $\chi^{(1)}(3\vec  q)=-1/\widetilde{\mathcal{K}}_{\rm tot}^{(2)}(3\vec  q)$,  we derive:

\begin{align}
    \rho(3\vec q)
=& \chi^{(1)}(3\vec  q)\left[ \chi^{(1)}(\vec  q)\right ]^3 \Gamma^{(3)}_{\rm tot}(\vec q)\, A^3, \label{eq:chi_cub_inter2}
\end{align}
where
\begin{align}
   \Gamma^{(3)}_{\rm tot}(\vec q)=&\left[
 \widetilde{\mathcal{K}}_{\rm tot}^{(3)}(3\vec q|\,2\vec q,\vec q)\,
\widetilde{\mathcal{K}}_{\rm tot}^{(3)}(2\vec  q|\,\vec  q,\vec  q)
\chi^{(1)}(2\vec  q)\right. \nonumber \\
& \quad \left.+  \widetilde{\mathcal{K}}_{\rm tot}^{(4)}(3\vec q|\,\vec q,\vec q,\vec q)
\right].
\end{align}
Finally, for the cubic response at third harmonic,  Eq.~(\ref{eq:chi_cub_inter}) then gives:
\begin{align}
    \chi^{(3)}(\vec q)
\equiv\,& \frac{\rho(3\vec q)}{A^3}=\chi^{(1)}(3\vec  q)\left[ \chi^{(1)}(\vec  q)\right ]^3  \Gamma^{(3)}_{\rm tot}(\vec q)\label{eq:chi_cub_3}\ .
\end{align}

Eq.~(\ref{eq:chi_cub_3}) is a general solution for the cubic response function at the third harmonic of electrons in homogeneous systems.   Let us now consider the UEG limit.

For the ideal UEG, Mikhailov expressed the ideal cubic   response function at the third harmonic $\chi^{(3)}_0(\mathbf{q})$ in terms of $\chi_0^{(1)}(\vec q)$~\cite{Mikhailov_Annalen,Mikhailov_PRL}:
\begin{equation}\label{eq:Mikhailov3}
\chi_0^{(3)}(\vec q)=\frac{3\chi_0^{(1)}(3\vec q)-8\chi^{(1)}_0(2\vec q)+5\chi_0^{(1)}(\vec q)}{3 \vec q^4}.
\end{equation}

For the ideal UEG, from Eq.~(\ref{eq:chi_cub_3}) and Eq.~(\ref{eq:Mikhailov3}), we find:
\begin{align}
&\frac{\chi_0^{(3)}(\vec q)}{\chi_0^{(1)}(3\vec  q)\left[ \chi_0^{(1)}(\vec  q)\right ]^3 }= \Gamma^{(3)}_s(\vec q),
\label{eq:gamma_s}
\end{align}
where $\Gamma^{(3)}_s(\vec q)$  follows from $\Gamma^{(3)}_{\rm tot}(\vec q)$ if we neglect the Hartree and XC terms:
\begin{align}
    \Gamma_s^{(3)}(\vec q)=&\,
 \widetilde{\mathcal{K}}_{s}^{(3)}(3\vec q|\,2\vec q,\vec q)\,
\widetilde{\mathcal{K}}_{s}^{(3)}(2\vec  q|\,\vec  q,\vec  q)
\chi_0^{(1)}(2\vec  q)\nonumber \\
&+  \widetilde{\mathcal{K}}_{s}^{(4)}(3\vec q|\,\vec q,\vec q,\vec q).
\end{align}

Now, considering the density response of the  interacting electrons with approximation $\Gamma^{(3)}_{\rm tot}(\vec q)\simeq \Gamma^{(3)}_s(\vec q)$ in Eq.~(\ref{eq:chi_cub_3}), using Eq.~(\ref{eq:gamma_s}) and Eq.~(\ref{eq:lind}), we deduce the result
 \begin{align}
     \chi^{(3)}( \vec q) =&\,  \chi^{(3)}_{0}( \vec q) \left[1-v(\vec q)\left[1-G(\vec q)\right]\chi^{(1)}_{0}(\vec q)\right]^{-3}\nonumber \\
     &\times \left[1-v(3\vec q)\left[1-G(3\vec q)\right]\chi^{(1)}_{0}( 3\vec q)\right]^{-1},\label{eq:chi3_LFC}
 \end{align}
where we used  $G(\vec q) = -\widetilde{\mathcal{K}}_{\rm xc}^{(2)}(\vec q) / v_c(\vec q)$.

In the RPA, setting $\widetilde{\mathcal{K}}_{\rm xc}^{(3)}(3\vec  q,2\vec  q,\vec  q)=0$,
$\widetilde{\mathcal{K}}_{\rm xc}^{(3)}(2\vec  q|\,\vec  q,\vec  q)=0$, $\widetilde{\mathcal{K}}_{\rm xc}^{(4)}(3\vec  q|\,\vec  q,\vec  q ,\vec  q)=0$, and $\widetilde{\mathcal{K}}_{\rm xc}^{(2)}(\vec q)=0$, we find from Eq.~(\ref{eq:chi_cub_3}):
\begin{align}
    \chi_{\rm RPA}^{(3)}(\vec q)
= \chi_{\rm RPA}^{(1)}(3\vec  q)\left[ \chi_{\rm RPA}^{(1)}(\vec  q)\right ]^3  \Gamma^{(3)}_{\rm RPA}(\vec q). \label{eq:chi_cub_3_rpa}
\end{align}

If we use the approximation $\Gamma^{(3)}_{\rm RPA}(\vec q)\simeq \Gamma^{(3)}_s(\vec q)$ in Eq.~(\ref{eq:chi_cub_3_rpa}), we derive the result from Ref.~\cite{Dornheim_PRR_2021}:

 \begin{align}
     \chi_{\rm RPA}^{(3)}( \vec q) =&\,  \chi^{(3)}_{0}( \vec q) \left[1-v(\vec q)\chi^{(1)}_{0}(\vec q)\right]^{-3}\nonumber \\
     &\times \left[1-v(3\vec q)\chi^{(1)}_{0}( 3\vec q)\right]^{-1}.
     \label{eq:chi3_RPA}
 \end{align}

The approximations for $\chi^{(3)}( \vec q) $ in the form of Eq.~(\ref{eq:chi3_LFC}) and Eq.~(\ref{eq:chi3_RPA}) were first presented in Ref.~\cite{Dornheim_PRR_2021}.
The presented derivations provide clarification regarding the approximations entering into Eq.~(\ref{eq:chi3_LFC}) and Eq.~(\ref{eq:chi3_RPA}). 
 Eq.~(\ref{eq:chi3_LFC}) and Eq.~(\ref{eq:chi3_RPA}) are based on the approximations   $\Gamma^{(3)}_{\rm tot}(\vec q)\simeq \Gamma^{(3)}_s(\vec q)$ and $\Gamma^{(3)}_{\rm RPA}(\vec q)\simeq \Gamma^{(3)}_s(\vec q)$, respectively.


\begin{figure}[!t]
\subfigure
{\includegraphics[width=0.52\linewidth]{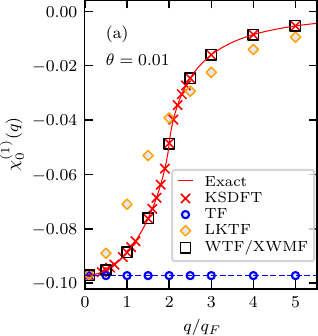}}
\subfigure
{\includegraphics[width= 0.465\linewidth]{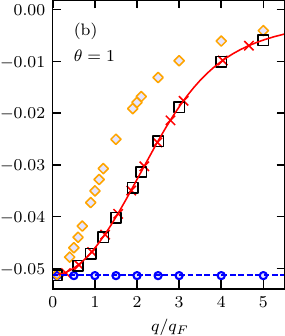}}
\caption{\label{fig:chi1_0} Linear static density response function of the ideal UEG at (a) $\theta=0.01$ and (b) $\theta=1$ with the density parameter set to $r_s=2$, and the wavenumber given in units of the Fermi wavenumber $q_F$.
The solid line is the Lindhard function, while the dashed line corresponds to the analytic solution for the TF model. Symbols indicate data computed using OFDFT with an external harmonic perturbation and various non-interacting free energy functionals. 
Note that the results from the WTF and XWMF for $\chi_0^{(1)}(\vec{q})$ are almost identical.}
\end{figure}

\section{Applications}\label{s:appl}

In this section, we demonstrate the application of the presented DFT framework by analyzing various non-interacting free energy functionals commonly used in orbital-free DFT (OFDFT). We specifically examine the Wang-Teter (WTF) functional \cite{Wang_Teter, WTF}, the LKTF functional \cite{LKTF, LKT_T0}, and the XWMF functional \cite{Cheng_PRB_2024, XWM_T0}. To facilitate this analysis, we have implemented the LKTF, WTF, and XWMF non-interacting free energy functionals into the open-source code DFTpy \cite{DFTpy}.

Additionally, we conducted Kohn-Sham DFT (KSDFT) calculations to generate exact data for the ideal non-linear density response functions. We consider the metallic electron density with a density parameter of $r_s = 2 $, where  $r_s$ represents the mean inter-electron distance in units of the Bohr radius~\cite{Ott2018}. This density is typical for metals and  warm dense matter applications.  We also vary the degeneracy parameter $\theta = k_\textnormal{B}T/E_F$, where $E_F$ is the Fermi energy, over the range from $\theta = 0.01$ to $\theta = 2$.

\subsection{Details of the simulations}

The OFDFT and KSDFT calculations of the ideal UEG perturbed by the external harmonic perturbation (\ref{eq:pert}) were performed using DFTpy and Quantum ESPRESSO \cite{Giannozzi_2009, Giannozzi_2017, Giannozzi_jcp_2020, Carnimeo_JCTC_2023}. To introduce the external harmonic perturbation in Quantum ESPRESSO calculations, we used QEpy \cite{QEpy}, which is a Python DFT engine for Quantum ESPRESSO-based nonstandard workflows. All calculations were performed using periodic boundary conditions, with the magnitude of the perturbation wavevector set to $q = i \times 2\pi / L_x$ (where $i$ is an integer), ensuring commensurability with the simulation box length along the direction of the perturbation wavevector.

In both OFDFT and KSDFT calculations, we utilized a real-space density grid with a spacing of $0.2 \, \text{Bohr}$. The OFDFT calculations were conducted with $118$ electrons in a cuboid with dimensions $L_y = L_z = 7.77 \, \text{Bohr}$ and $L_x = 65.49 \, \text{Bohr}$. The same simulation cell containing $118$ electrons was used for the KSDFT calculations at $\theta = 0.01$ with the number of bands set to $N_b = 80$, and using a $2 \times 60 \times 60$ k-point grid. For higher temperatures ($\theta = 0.25$, $\theta = 0.5$, $\theta = 1$, and $\theta = 2$), the KSDFT calculations were performed with $N = 38$ electrons in a cuboid with side lengths $L_y = L_z = 7.77 \, \text{Bohr}$ and $L_x = 21.09 \, \text{Bohr}$. The k-point grid for these simulations was set to $4 \times 16 \times 16$.  The number of bands was adjusted according to the temperature: $N_b = 140$ at $\theta = 0.25$, $N_b = 280$ at $\theta = 0.5$, $N_b = 600$ at $\theta = 1$, and $N_b = 1400$ at $\theta = 2$. In all KSDFT calculations, we applied a cutoff energy of $62 \, \text{Ry}$.  
The input files used for the KSDFT and OFDFT calculations are available on the Rossendorf Data Repository \cite{moldabekov_zhandos_2026_4487}.

\subsection{Results of the calculations}
\subsubsection{Linear density response  $\chi_0^{(1)}(\vec{q})$}
To begin with, we consider the quality of $\chi_0^{(1)}(\vec{q})$ computed using different approximations for $F_s[n]$, as well as how $\chi_0^{(1)}(\vec{q})$ depends on wavenumber and temperature. Understanding these aspects, as we will see, is crucial for interpreting the results for the ideal non-linear density response functions. 

In Fig.~\ref{fig:chi1_0}, for $\theta=0.01$ and $\theta=1$, we compare the OFDFT and KSDFT results for $\chi_0^{(1)}(\vec{q})$ with the Lindhard function (which is the exact solution for $\chi_0^{(1)}(\vec{q})$ of the ideal UEG). As expected, the KSDFT data show excellent agreement with the Lindhard function. The OFDFT calculations using the WTF and XWMF also reproduce the Lindhard function, as these functionals are designed to reduce to it in the UEG limit.  The dependence of the GGA level LKTF result for the response function $\chi_0^{(1)}(\vec{q})$ on the wavenumber is also as anticipated \cite{LKTF, LKT_T0}. The LKTF result tends toward the TF limit at small wavenumbers and exhibits a decreasing magnitude as the wavenumber increases, with the quality aligning with the expectations from the $F_s[n]$ with a gradient correction \cite{Jones_1971, Kirzhnits_1975, Moldabekov_pop_2018, moldabekov_pop_15, Akbari}. The TF result for $\chi_0^{(1)}(\vec{q})$ agrees with the exact result in the limit $\vec{q} \to 0$ and remains constant regardless of the wavenumber of the perturbation. In addition, in Fig.~\ref{fig:chi1_0}, we see that the results from the TF model-based OFDFT simulations of the harmonically perturbed ideal electron gas (circles) are in agreement with the theoretical result using the second-order functional derivative of the TF functional (dashed horizontal lines). The latter is provided by Eq.~(\ref{eq:ktf2}).

\begin{figure}[!t]
\subfigure
{\includegraphics[width=1\linewidth]{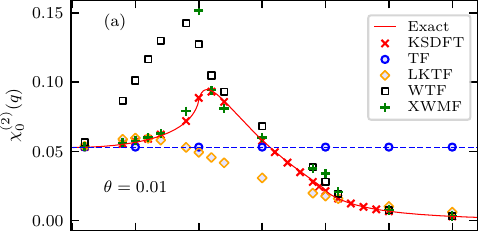}}\\
\vspace{-1.0\baselineskip}
\subfigure
{\includegraphics[width= 1\linewidth]{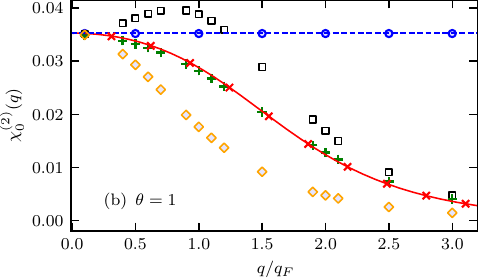}}
\caption{\label{fig:chi2_0} Quadratic static density response function of the ideal UEG at (a) \(\theta=0.01\) and (b) \(\theta=1\) with the density parameter \(r_s=2\). The solid line represents the exact solution given by Eq. \eqref{eq:chi2_0} and the dashed line shows the analytic solution using the TF model. Symbols correspond to the OFDFT calculations using the external harmonic perturbation and different approximations for the non-interacting free energy functional.
 }
\end{figure}

\subsubsection{Quadratic density response  $\chi_0^{(2)}(\vec{q})$}

In Fig.~\ref{fig:chi2_0}, we present the results for the quadratic density response at the second harmonic, $\chi_0^{(2)}(\vec{q})$, at $\theta=0.01$ and $\theta=1$. We compare the results obtained from OFDFT and KSDFT with the exact solution (\ref{eq:chi2_0}) given by Mikhailov~\cite{mikhailov2012}. As it should be, the KSDFT data agree with the analytical solution (\ref{eq:chi2_0}). 
From Fig.~\ref{fig:chi2_0}(a), we observe that the exact solution for $\chi_0^{(2)}(\vec{q})$ exhibits a positive peak at approximately $q \simeq q_F$. The origin of this maximum can be understood by recalling that $\chi_0^{(2)}(\vec{q})$ is proportional to the difference $\chi^{(1)}_0(2\vec{q}) - \chi^{(1)}_0(\vec{q})$ (see Eq.~\ref{eq:chi2_0}). At $\theta=0.01$, the slope of $\chi^{(1)}_0(\vec q)$ has the largest value around $q=2q_F$ (as one can see from Fig.~\ref{fig:chi1_0}(a)), while its $q$-dependence is much weaker for $q\ll q_F$ and $q\gtrsim 3q_F$. As a result, the difference $\chi^{(1)}_0(2\vec{q}) - \chi^{(1)}_0(\vec{q})$ increases most rapidly around $q \simeq q_F$, producing the pronounced peak observed in Fig.~\ref{fig:chi2_0}(a).
As the temperature increases to $\theta = 1$, this peak disappears, as illustrated in Fig.~\ref{fig:chi2_0}(b). This results from the thermally induced reduction of the slope of  $\chi^{(1)}_0(\vec{q})$ at $q = 2q_F$ (see Fig.~\ref{fig:chi1_0}(b)).

\begin{figure}[!t]
{\includegraphics[width=1\linewidth]{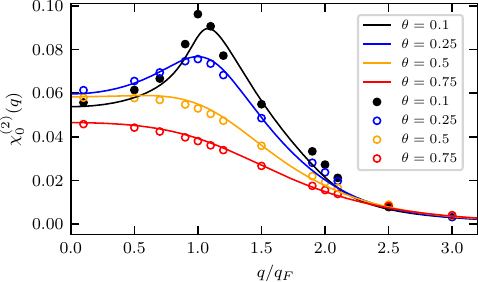}}
\caption{\label{fig:chi2_0_theta} Quadratic static density response function at second harmonic of the ideal UEG at different values of the degeneracy parameter with the density parameter set to \(r_s=2\). The lines correspond to the exact solution given by Eq. \eqref{eq:chi2_0}. Circles depict the results of the OFDFT calculations using the XWMF non-interacting free-energy functional.
 }
\end{figure}

From Fig.~\ref{fig:chi2_0}, we see that for both $\theta = 0.01$ and $\theta = 1$,  the results from the LKTF-based OFDFT calculations for $\chi_0^{(2)}(\vec{q})$ align well with the exact solution at the limits of small and large wavenumbers. However, at intermediate wavenumbers where $0.1\,q_F \lesssim q \lesssim 2\,q_F$, the LKTF data exhibits significant disagreement with the exact solution. While this behavior is expected for the LKTF, it is surprising to see that the WTF also performs poorly in describing $\chi_0^{(2)}(\vec{q})$.  The WTF functional is specifically designed to reproduce $\chi_0^{(1)}(\vec{q})$, which is supported by simulations of the perturbed ideal UEG, as shown in Fig.~\ref{fig:chi1_0}. In principle, this should suffice for providing an accurate description of $\chi_0^{(2)}(\vec{q})$, as indicated by Eq.~(\ref{eq:chi2_0}). However, Fig.~\ref{fig:chi2_0} reveals that this is not the case for the WTF. This discrepancy clearly suggests that the ansatz used to construct the WTF functional violates the exact constraint given in Eq.~(\ref{eq:Ks3_chi0}). In contrast, XWMF, a functional developed using the method of line integrals and designed to reproduce the $\chi_0^{(1)}(\vec{q})$ of the ideal UEG, shows close agreement with the exact data for $\chi_0^{(2)}(\vec{q})$. The only exceptions are in the regions around $q = q_F$ and $q = 2q_F$ at $\theta = 0.01$. This behavior of XWMF is attributed to a numerical issue that arises when computing the ground state $\chi^{(1)}_0(\vec{q})$ at $q = 2q_F$. 
This issue stems from the term $f(z) = (1-z^2)\ln\left|\frac{1+z}{1-z}\right|$ (where $z = \frac{q}{2q_F}$) in the Lindhard function \cite{Lindhard1954, Arista_pra_1984}. Although it is true that $\lim_{z \to 0} f(z) \to 0$ theoretically, numerically one has to use a finite value for the logarithm in $f(z)$ in the vicinity of $z = 1$. This numerical peculiarity has a negligible effect on $\chi_0^{(1)}(\vec{q})$, but it can lead to significant inaccuracies in $\chi_0^{(2)}(\vec{q})$ around $q = q_F$ and $q = 2q_F$, if not carefully treated. This is because $\chi_0^{(2)}(\vec{q})$ is proportional to the difference between $\chi_0^{(1)}(2\vec{q})$ and $\chi_0^{(1)}(\vec{q})$ (see Eq.~(\ref{eq:chi2_0})), which makes it more sensitive to the accuracy of the calculations. This issue diminishes at $\theta=1$. At this condition with strong thermal excitations, $\chi_0^{(2)}(\vec{q})$ does not exhibit noticeable non-monotonic behavior (see Fig.~\ref{fig:chi2_0}(b)).

\begin{figure}[!t]
\subfigure{
\hspace{+0.03cm}
{\includegraphics[width=0.98\linewidth]{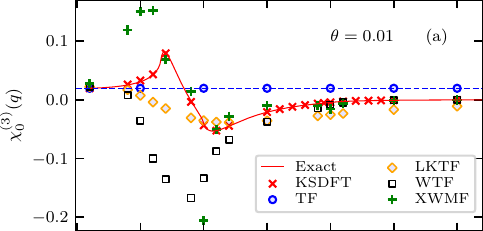}}}\\
\vspace*{-1.0\baselineskip}
\subfigure{
{\includegraphics[width=0.48\textwidth]{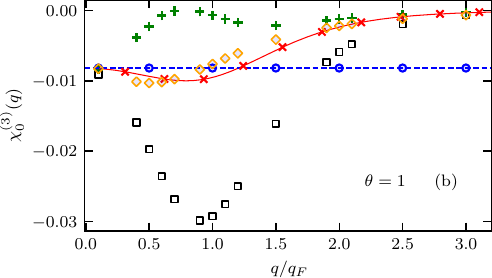}}}
\caption{\label{fig:chi3_0} Cubic static density response function of the ideal UEG at (a) \(\theta=0.01\) and (b) \(\theta=1\) with the density parameter \(r_s=2\). The solid line represents the exact solution given by Eq. \eqref{eq:Mikhailov3} and the dashed line is the analytic solution using the TF model. Symbols correspond to the OFDFT calculations using the external harmonic perturbation and different approximations for the non-interacting free energy functional.
 }
\end{figure}

In Fig.~\ref{fig:chi2_0_theta}, we present $\chi_0^{(2)}(\vec{q})$ at various values of $\theta$ to illustrate the correlation between the enhanced performance of the XWMF and the suppression of the maxima of $\chi_0^{(2)}(\vec{q})$ due to thermal excitations. From Fig.~\ref{fig:chi2_0_theta}, it is evident that for $\theta \geq 0.25$, the XWMF results have a good agreement with the exact solution for $\chi_0^{(2)}(\vec{q})$ without significant issues around $q = q_F$ and $q = 2q_F$. Thus, in terms of describing the quadratic response function, the XWMF demonstrates superior performance under WDM conditions compared to other non-interacting free energy functionals that have been considered.

\subsubsection{Cubic density response at third harmonic  $\chi_0^{(3)}(\vec{q})$}

In Fig.~\ref{fig:chi3_0}, the results for $\chi_0^{(3)}(\vec{q})$ at $\theta = 0.01$ and $\theta = 1$ are presented. As expected, the KSDFT calculations closely match the exact analytic solution in equation (\ref{eq:Mikhailov3}). At $\theta = 0.01$, $\chi_0^{(3)}(\vec{q})$ displays a pronounced positive peak around $q = \frac{2}{3}q_F$ and a negative minimum at $q = q_F$. The origin of this structure can be explained by looking at the recursive relation in equation (\ref{eq:Mikhailov3}). From this relation, we can see that $\chi_0^{(3)}(\vec{q})$ is determined by the combination $3\chi_0^{(1)}(3\vec{q}) - 8\chi_0^{(1)}(2\vec{q}) + 5\chi_0^{(1)}(\vec{q})$ of the ideal linear density response functions. This, combined with the sharp slope of $\chi_0^{(1)}(\vec{k})$ around $k = 2q_F$, results in a strong non-monotonic dependence of $\chi_0^{(3)}(\vec{q})$ on the wavenumber near $q = \frac{2}{3}q_F$ and $q = q_F$.

\begin{figure}[!t]
{\includegraphics[width=1\linewidth]{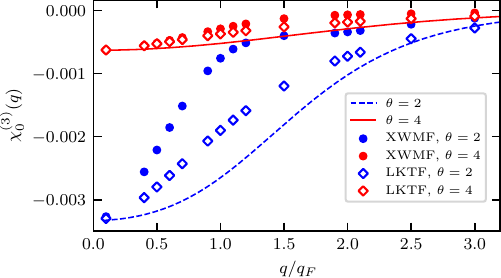}}
\caption{\label{fig:chi3_0_theta} Cubic static density response function at third harmonic of the ideal UEG at $\theta=2$ and $\theta=4$ with $r_s=2$. The lines correspond to the exact solution given by Eq. \eqref{eq:Mikhailov3}. Symbols show the results of the OFDFT calculations using the XWMF and LKTF non-interacting free-energy functionals.
 }
\end{figure}

From Fig.~\ref{fig:chi3_0}, we can confirm that the TF result accurately provides the $q \to 0$ limit for $\chi_0^{(3)}(\vec{q})$. This is supported by the agreement between the analytic solution (depicted as horizontal dashed lines) and the simulation data for the perturbed electron gas (represented by circles). The analytic solution for the TF model is given by Eqs.~(\ref{eq:ktf3}) and (\ref{eq:chi30TF}). The data from the WTF-based OFDFT calculations for $\chi_0^{(3)}(\vec{q})$ align with the exact solution at both small and large wavenumbers. However, significant deviations are observed at intermediate wavenumbers, specifically in the range of $0.1\,q_F \lesssim q \lesssim 2\,q_F$. In comparison to the WTF, the LKTF model provides a much more accurate description of $\chi_0^{(3)}(\vec{q})$. Although LKTF is based on GGA, which, as expected, does not capture a sharp positive peak for $\chi_0^{(3)}(\vec{q})$ around $  q\simeq 2/3\,q_F$, it does manage to identify a pattern with a negative minimum. Furthermore, Fig.~\ref{fig:chi3_0} indicates that the XWMF performs worse for $\chi_0^{(3)}(\vec{q})$ compared to the cases of the ideal quadratic and linear density response functions. In fact, we can assess that LKTF is more accurate than XWMF for $\chi_0^{(3)}(\vec{q})$, particularly at $\theta=1$ (as shown in Fig.~\ref{fig:chi3_0}(b)). This trend persists for larger values of $\theta$, as illustrated in Fig.~\ref{fig:chi3_0_theta}, which presents results for $\chi_0^{(3)}(\vec{q})$ computed at $\theta=2$ and $\theta=4$ using both LKTF and XWMF. The decline in the performance of the XWMF for $\chi_0^{(3)}(\vec{q})$, relative to the ideal quadratic and linear response functions, is likely due to the peculiarities of the line integrals used.

As a third-order contribution with respect to the magnitude of the perturbation, the poor performance of the XWMF and WTF for $\chi_0^{(3)}(\vec{q})$ does not allow for an adequate analysis of $\chi_0^{(1, 3)}(\vec{q})$ when employing these functionals in simulations of the harmonically perturbed UEG.

\begin{figure}[!t]
\subfigure{
\hspace{+0.05cm}
{\includegraphics[width=0.975\linewidth]{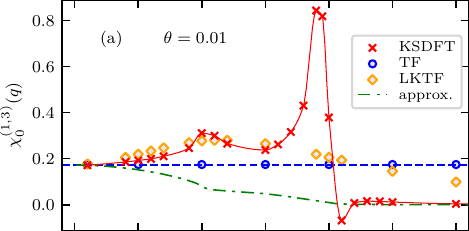}}}\\
\vspace*{-1.0\baselineskip}
\subfigure{
{\includegraphics[width=0.479\textwidth]{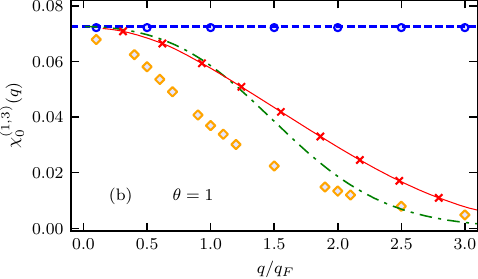}}}
\caption{\label{fig:chi13_0} Cubic static density response function of the ideal UEG at the first harmonic for (a) $\theta = 0.01$ and (b) $\theta = 1$, with density parameter $r_s = 2$.  The KSDFT results are shown as cross markers, with an interpolated line included to guide the eye. 
The horizontal dashed line denotes the analytic Thomas--Fermi (TF) solution. 
Other symbols represent OFDFT calculations using different approximations to the non-interacting free-energy functional. The dash-dotted line depicts analytical approximation (\ref{eq:chi13_app}).
 }
\end{figure}

\begin{figure}[htbp!]
{\includegraphics[width=1\linewidth]{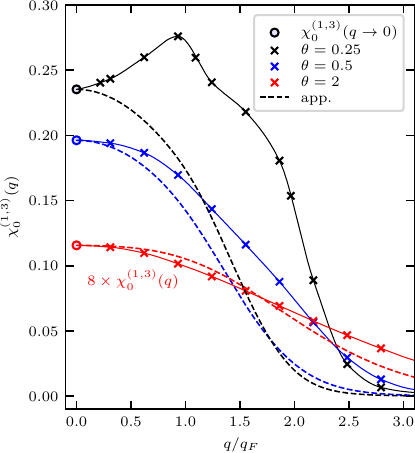}}
\caption{\label{fig:chi13_theta} Cubic static density response function at first harmonic of the ideal UEG at different values of the degeneracy parameters and $r_s=2$. The KSDFT results are depicted using markers, with an interpolated line included to guide the eye. The dashed line corresponds to approximation (\ref{eq:chi13_app}).
 }
\end{figure}

\begin{figure}[htbp]
{\includegraphics[width=0.95\linewidth]{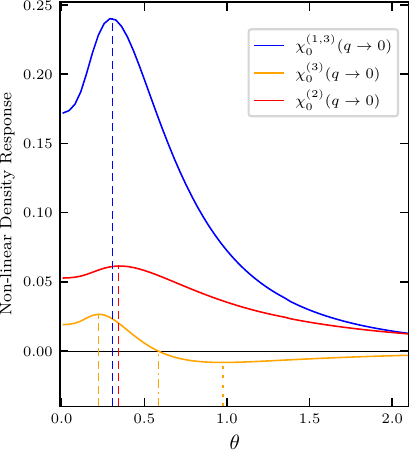}}
\caption{\label{fig:chi13_q0} The dependence of the long-wavelength limit of the ideal non-linear response functions of the UEG on the degeneracy parameter $\theta$. The dashed vertical lines indicate the positions of the maxima of \(\chi_0^{(1,3)}(q\to0)\), \(\chi_0^{(2)}(q\to0)\), and \(\chi_0^{(3)}(q\to0)\). The dotted line represents the minimum of \(\chi_0^{(3)}(q\to0)\), while the dash-dotted line indicates the zero value of \(\chi_0^{(3)}(q\to0)\). The function \(\chi_0^{(1,3)}(q\to0)\) reaches its maximum value at \(\theta \simeq 0.308\), and \(\chi_0^{(2)}(q\to0)\) reaches its maximum at \(\theta \simeq 0.343\). Additionally, \(\chi_0^{(3)}(q\to0)\) has a maximum value at \(\theta \simeq 0.22\), a minimum value at \(\theta \simeq  0.975\), and crosses zero at \(\theta \simeq 0.58\).
 }
\end{figure}

\subsubsection{Cubic density response at first harmonic  $\chi_0^{(1,3)}(\vec{q})$}

In Figure~\ref{fig:chi13_0}, we present the KSDFT results for $\chi_0^{(1,3)}(\vec{q})$ at two temperatures: $\theta = 0.01$ and $\theta = 1$. From Figure~\ref{fig:chi13_0}(a), it is clear that $\chi_0^{(1,3)}(\vec{q})$ exhibits a pronounced non-monotonic dependence on the perturbation wavenumber. Specifically, it features two peaks: one at approximately $q \simeq q_F$ and another at $q \simeq 1.85 \, q_F$, along with a negative minimum at $q \simeq 2.1 \, q_F$. Additionally, there is a notable drop from positive to negative values around $q = 2 q_F$.
This non-monotonic behavior is similar to what is observed in the second- and third-order response functions, $\chi_0^{(2)}(\vec{q})$ and $\chi_0^{(3)}(\vec{q})$, at $\theta=0.01$. For $\chi_0^{(1,3)}(\vec{q})$, the structure around $q = q_F$ and $q = 2, q_F$ can be traced back to the aforementioned peculiar feature of the linear response function $\chi_0^{(1)}(\vec{q})$ near twice the Fermi wavenumber, where the maximum slope (i.e., the maximum magnitude of its derivative) is located. As established in Eq.~(\ref{eq:chi13_new2}), $\chi_0^{(1,3)}(\vec{q})$ arises from mode coupling between density perturbations at $\vec{k} = \vec{q}$ and $\vec{k} = 2\vec{q}$. Consequently, strong non-monotonic features appear for perturbation wavenumbers at the Fermi wavenumber and its second harmonic.

The non-monotonic behavior of $\chi_0^{(1,3)}(\vec{q})$  disappears entirely as the temperature increases to $\theta = 1$, as shown in Figure~\ref{fig:chi13_0}(b). At this higher temperature, $\chi_0^{(1,3)}(\vec{q})$ becomes positive and decreases monotonically with increasing wavenumber.

From Fig.~\ref{fig:chi13_0}, we observe that for both temperatures, $\theta = 0.01$ and $\theta = 1$, the KSDFT data align with the TF  results at small wavenumbers. Furthermore, we observe agreement between the exact analytical result (\ref{eq:chi0130}) (also see (\ref{eq:chi130TF})) for the $\vec{q} \to 0$ limit of $\chi_0^{(1,3)}(\vec{q})$, derived using the TF model (indicated by the horizontal dashed line), and the TF-based OFDFT simulations of the perturbed ideal UEG  (shown as circles). At $\theta = 0.01$, the OFDFT results obtained using the LKFT functional provide a satisfactory description of $\chi_0^{(1,3)}(\vec{q})$ for $q \lesssim 1.5 q_F$. However, the LKTF does not capture the strong non-monotonic features of $\chi_0^{(1,3)}(\vec{q})$ around $q = 2 \, q_F$, which is beyond the capabilities of the GGA level description. At $\theta = 1$, the LKTF also demonstrates a monotonically decreasing trend with increasing wavenumber. While it is quantitatively accurate at both small and large wavenumbers, the LKTF shows disagreement with the KSDFT data in the intermediate range. 

In Fig.~\ref{fig:chi13_0}, we also compare approximation (\ref{eq:chi13_app}) with the KSDFT data. It can be seen that Eq.~(\ref{eq:chi13_app}) (dash-dotted line) does not capture the non-monotonic behavior of $\chi_0^{(1,3)}(\vec{q})$ at $\theta=0.01$, and it provides quantitative agreement with the KSDFT results only in the limit of small wavenumbers, where $\chi_0^{(1,3)}(\vec{q})$ values are close to the TF data. At $\theta=1$, the approximation (\ref{eq:chi13_app}) adequately captures the monotonically decreasing pattern of $\chi_0^{(1,3)}(\vec{q})$ and yields quantitatively accurate results for $q \lesssim q_F$.

In Fig.~\ref{fig:chi13_theta}, we present KSDFT data for $\chi_0^{(1,3)}(\vec{q})$ at $\theta=0.25$, $\theta=0.5$, and $\theta=2$. The values of $\chi_0^{(1,3)}(\vec{q})$ at the limit $\vec{q} \to 0$ are computed using the derived exact analytical result (\ref{eq:chi0130}). We clearly see the thermal excitations induced suppression of the non-monotonic behavior of $\chi_0^{(1,3)}(\vec{q})$. At $\theta=0.25$, $\chi_0^{(1,3)}(\vec{q})$ exhibits only one maximum and does not reach any negative values. Further increasing $\theta$ to $\theta=0.5$ leads to a monotonically decreasing behavior of $\chi_0^{(1,3)}(\vec{q})$ as the wavenumber increases. Moreover, the rate of decrease of $\chi_0^{(1,3)}(\vec{q})$ with increasing wavenumber slows down as $\theta$ increases, as illustrated by the result for $\theta=2$ in Fig.~\ref{fig:chi13_theta} and $\theta=1$ in Fig.~\ref{fig:chi13_0}. As $\theta$ increases, the quality of approximation (\ref{eq:chi13_app}) improves, leading to a better quantitative description of $\chi_0^{(1,3)}(\vec{q})$.

In Fig.~\ref{fig:chi13_q0}, we illustrate how the $\vec{q} \to 0$ limits of the examined non-linear density response functions depend on the degeneracy parameter $\theta$. As discussed in Appendix~\ref{s:TF_kernels}, the small-$\vec{q}$ limits of $\chi_0^{(2)}(\vec{q})$, $\chi_0^{(3)}(\vec{q})$, and $\chi_0^{(1,3)}(\vec{q})$ depend solely on $\theta$. From Fig.~\ref{fig:chi13_q0}, it is evident that these limits exhibit a non-monotonic dependence on $\theta$: all three functions initially increase with increasing $\theta$, reach a positive maximum, and then decrease as $\theta$ continues to rise.
Quantitatively, $\chi_0^{(1,3)}(q \to 0)$ attains its maximum at $\theta \simeq 0.308$, while $\chi_0^{(2)}(q \to 0)$ reaches its maximum at $\theta \simeq 0.343$. The cubic response at the third harmonic, $\chi_0^{(3)}(q \to 0)$, shows a maximum at $\theta \simeq 0.22$, a minimum at $\theta \simeq 0.975$, and crosses zero near $\theta \simeq 0.58$.
These trends indicate that non-linear electronic response effects are most pronounced for partially degenerate electrons with $\theta <0.5$.

\section{Conclusions and Outlook}

We investigated the relationship between static non-linear density response functions and the functional derivatives of free-energy functionals. For systems with an average homogeneous density distribution, Eq.~(\ref{eq:master_compact2}) provides a general DFT-based framework for describing the non-linear response of electrons. This framework allowed us to uncover the underlying mode-coupling structure responsible for the previously unresolved cubic response at the first harmonic. For the ideal UEG, it also yields exact analytical long-wavelength limits for all non-linear response functions considered. The resulting connection between $F_s[n]$ and the ideal non-linear response functions provides useful constraints for constructing non-interacting free-energy functionals.

As a practical application, we emphasize that a key element in developing approximations to the non-interacting free-energy functionals for metals, semiconductors, and warm dense matter is the link between the ideal linear density response function of the UEG and the second-order functional derivative of $F_s[n]$, given by Eq.~\eqref{eq:lindhard}. At low temperatures ($T \to 0$), $F_s[n]$ reduces to the kinetic-energy functional. In a similar way, Eqs.~\eqref{eq:Ks3_chi0} and \eqref{eq:gamma_s}, which connect the third- and fourth-order functional derivatives of $F_s[n]$ to $\chi_0^{(2)}(\vec{q})$ and $\chi_0^{(3)}(\vec{q})$, provide additional exact constraints. These constraints can be useful when selecting or validating an ansatz for constructing a particular approximation for $F_s[n]$. This is motivated by the fact that the quadratic density response at the second harmonic, $\chi_0^{(2)}(\vec{q})$, and the cubic response at the third harmonic, $\chi_0^{(3)}(\vec{q})$, can be expressed as combinations of ideal linear density response functions~\cite{Mikhailov_Annalen, Mikhailov_PRL} (see Eqs.~\eqref{eq:chi2_0} and \eqref{eq:Mikhailov3}).

Our OFDFT simulations show that the WTF functional, although it reproduces $\chi^{(1)}_0(\vec{q})$, violates the exact quadratic-response relation and therefore fails to describe $\chi^{(2)}_0(\vec{q})$ across the relevant wave-number range. In contrast, the XWMF functional provides a reasonable description of the quadratic response. Both the WTF and XWMF functionals, however, fail to capture the cubic density response at the first and third harmonics. The LKTF functional performs more robustly, although it cannot reproduce the sharp non-monotonic features in the ideal non-linear response that arise from the steep slope of the Lindhard function near $2 q_F$ in the strongly degenerate regime ($\theta \ll 1$). Approximations to $F_s[n]$ that accurately describe non-linear density responses are expected to be important in the warm dense matter regime, at parameters where an accurate treatment of electron screening of ion-ion pair interaction potential requires going beyond the linear-response regime~\cite{Ashcroft_prl_2003, Ashcroft_PRB_2007, moldabekov_cpp_22, zhandos_cpp17}. In this context, satisfying the quadratic-response constraint of Eq.~\eqref{eq:chi2_0} is particularly important, as it represents the next leading contribution after the linear response~\cite{Dornheim_PRR_2021}.

Using our framework, we have also obtained the first exact KSDFT data for $\chi_0^{(1,3)}(\vec{q})$ in the UEG. These results reveal strong non-monotonic features at low temperature that originate from mode coupling between perturbations at $\vec{q}$ and $2\vec{q}$. These structures gradually diminish with increasing temperature and become fully monotonic for $\theta \gtrsim 0.5$. We also find that the exact long-wavelength limits of all non-linear response functions exhibit a non-monotonic dependence on the degeneracy parameter, indicating that non-linear effects are strongest for partially degenerate electrons with $\theta<0.5$.

Overall, the present work establishes a systematic route for using non-linear response theory to inform the development, evaluation, and refinement of orbital-free DFT functionals. Future applications include constructing constrained non-interacting functionals consistent with the hierarchy of ideal response relations. Another promising direction is extending constraints for exchange--correlation functionals by incorporating exact results for the kernel $\widetilde{\mathcal{K}}_{\rm xc}^{(3)}(2\vec{q}\,|\,\vec{q},\vec{q})$ (see Eq.~(\ref{eq:kxc3})), which can be computed using various quantum Monte Carlo methods~\cite{review,Bonitz_pop_2024}. Finally, the framework can also be used to investigate time-dependent non-linear response in warm dense matter and quantum plasmas by analyzing the electronic response to perturbations of the form $\Delta v_{\rm ext} \propto \cos(\vec{q}\cdot\vec{r} - \omega t)$.

\section*{Data Availability}
The data supporting the findings of this study are available on the Rossendorf Data Repository (RODARE)~\cite{moldabekov_zhandos_2026_4487}.

\begin{acknowledgments}

This work was partially supported by the Center for Advanced Systems Understanding (CASUS), financed by Germany’s Federal Ministry of Education and Research (BMBF) and the Saxon state government out of the State budget approved by the Saxon State Parliament. This work has received funding from the European Research Council (ERC) under the European Union’s Horizon 2022 research and innovation programme
(Grant agreement No. 101076233, "PREXTREME"). 
Views and opinions expressed are however those of the authors only and do not necessarily reflect those of the European Union or the European Research Council Executive Agency. Neither the European Union nor the granting authority can be held responsible for them. Tobias Dornheim gratefully acknowledges funding from the Deutsche Forschungsgemeinschaft (DFG) via project DO 2670/1-1. 
This work has received funding from the German Federal Ministry of Research, Technology and Space (BMFTR) via the ErUM Data project "DEMOS" (05D25CR1).
Computations were performed on a Bull Cluster at the Center for Information Services and High-Performance Computing (ZIH) at Technische Universit\"at Dresden and at the Norddeutscher Verbund f\"ur Hoch- und H\"ochstleistungsrechnen (HLRN) under grant mvp00024.
\end{acknowledgments}

\appendix

\section{Thomas--Fermi reduced kernels}\label{s:TF_kernels}

For the TF potential $v_{\rm TF}[n]=\mu(\eta[n])$ \cite{Teller_1949}, the reduced kernels read:
\begin{align}
\label{eq:ktf2}
\widetilde{\mathcal{K}}^{(2)}_{\rm TF} &= \frac{1}{\beta}\,\frac{\lambda_T^3}{g}\frac{1}{F_{-1/2}(\eta)},   \\
\widetilde{\mathcal{K}}^{(3)}_{\rm TF} &= -\frac{1}{2}\,\frac{1}{\beta}\,\left(\frac{\lambda_T^3}{g}\right)^{\!2}
\,\frac{F_{-3/2}(\eta)}{\left[F_{-1/2}(\eta)\right]^3}, \label{eq:ktf3} \\
\widetilde{\mathcal{K}}^{(4)}_{\rm TF} &= \frac{1}{6}\,\frac{1}{\beta}\,\left(\frac{\lambda_T^3}{g}\right)^{\!3}
\,\frac{\,3\,\left[F_{-3/2}(\eta)\right]^2-F_{-1/2}(\eta)F_{-5/2}(\eta)\,}{\left[F_{-1/2}(\eta)\right]^5},\label{eq:ktf4}
\end{align}
where $ \lambda_T = \sqrt{2\pi \beta}$, $\beta=1/T$, $g=2$, $\eta=\mu\beta$,
and $F_{\nu}(\eta)$ is the Fermi integral of order $\nu$. We calculated the Fermi integrals using $F_{\nu}(\eta) = -{\rm Li}_{\nu+1}(-e^{x})$, with the polylogarithm computed using the Python mpmath library.

The chemical potential $\eta=\mu\beta$ is defined by the temperature and density through the relation:
\begin{equation}
    \frac{2}{3}\theta^{-3/2}=\int_0^{\infty}\frac{\sqrt{x}}{1+\exp(x-\eta)}~{\rm d}x\, ,
\end{equation}
with $\theta=T/E_F=2T/(3\pi^2n)^{2/3}$ being the degeneracy parameter.

Eq. (\ref{eq:ktf2}) defines the long wavelength limit of the ideal density response function $\chi_0^{(1)}(\vec q\to 0)=-1/\widetilde{\mathcal{K}}^{(2)}_{\rm TF}$.
Using Eq.~\eqref{eq:chi2_tot} and the TF reduced kernels, for the long wavelength limit of the ideal quadratic density response function, we find:

\begin{equation}
    \chi^{(2)}(\vec q \to 0)=-\frac{\widetilde{\mathcal{K}}_{\rm TF}^{(3)}}{\left[\widetilde{\mathcal{K}}^{(2)}_{\rm TF}\right]^3}.\label{eq:chi2_TF}
\end{equation}

Following Eq.~(\ref{eq:chi0130}), for the cubic response at first harmonic, we have:
\begin{equation}\label{eq:chi130TF}
     \chi_0^{(1,3)}(\vec q \to 0)=\frac{\Gamma_{\rm TF}^{(1,3)}}{\left[\widetilde{\mathcal{K}}^{(2)}_{\rm TF}\right]^{4} },
\end{equation}
where
\begin{align}
    \Gamma_{\rm TF}^{(1,3)}(\vec q)&=\,\widetilde{\mathcal{K}}_{\rm TF}^{(4)}-2\,\frac{\left[\widetilde{\mathcal{K}}_{\rm TF}^{(3)}\right]^2}{\widetilde{\mathcal{K}}^{(2)}_{\rm TF}}.
\end{align}

Using Eq.~(\ref{eq:chi_cub_3}) and the TF kernels, for the cubic response at third harmonic, we have:
\begin{equation}\label{eq:chi30TF}
     \chi_0^{(3)}(\vec q \to 0)=\frac{\Gamma_{\rm TF}^{(3)}}{\left[\widetilde{\mathcal{K}}^{(2)}_{\rm TF}\right]^{4} },
\end{equation}
where
\begin{align}
   \Gamma^{(3)}_{\rm TF}(\vec q)=&
 -\frac{\left[\widetilde{\mathcal{K}}_{\rm TF}^{(3)}\right]^2}{
\widetilde{\mathcal{K}}_{\rm TF}^{(2)}} +  \widetilde{\mathcal{K}}_{\rm TF}^{(4)}.
\end{align}
\\

\bibliography{bibliography}
\end{document}